\documentclass[12pt,preprint]{aastex}
\slugcomment{Accepted 2020 December 15, Received 2020 October 13 and scheduled for publication 
in  the  Monthly Notices}
\def\lax {\ifmmode{_<\atop^{\sim}}\else{${_<\atop^{\sim}}$}\fi}
\def\gax {\ifmmode{_>\atop^{\sim}}\else{${_>\atop^{\sim}}$}\fi}
\def\gtorder{\mathrel{\raise.3ex\hbox{$>$}\mkern-14mu
             \lower0.6ex\hbox{$\sim$}}}

\begin{document}

%
%

   \title
{Detection of a High-Temperature Blackbody  Hump in  Black Hole Spectra.
The strongly redshifted annihilation line}

\author{Lev Titarchuk\altaffilmark{1, 2} and Elena Seifina\altaffilmark{2,3} 
}

\altaffiltext{1}{
Astro Space Center, Lebedev Physical Institute, Russian Academy of Sciences,  Profsouznay ul. 84/32, Moscow 117997, Russia;
ltitarchuk@asc.rssi.ru; titarchuk@fe.infn.it
}

\altaffiltext{2}{Lomonosov Moscow State University/Sternberg Astronomical Institute,
Universitetsky Prospect 13, Moscow, 119992, Russia; seif@sai.msu.ru
}

\altaffiltext{3}{LAPTh, Universite Savoie Mont Blanc, CNRS, B.P. 110, Annecy-le-Vieux
F-74941, France; 
All-Russian Institute of Scientific and Technical Information, RAS, 
Usievich st. 20,  Moscow, 125190, Russia
}





\begin{abstract}
{We discovered a so called  high-temperature blackbody (HBB)  component, found 
in the 15 -- 40 keV range, in the  broad-band 
X-ray energy spectra of black hole (BH) candidate sources.  
A detailed study of this spectral feature is presented using  data from five of the  Galactic BH binaries, 
Cyg~X--1, GX~339--4, GRS~1915+105,  SS~433 and V4641~Sgr 
in the low/hard,  intermediate, high/soft and very soft spectral states  (LHS, IS, HSS and VSS, respectively) and spectral transitions between them using {\it RXTE}, INTEGRAL and {\it Beppo}SAX data.
 In order to fit the  broad-band energy spectra of these sources we used 
an additive  XSPEC model, composed of 
the  Comptonization 
component  and the Gaussian line component.
In particular, we reveal that  the IS spectra have the HBB component  which color temperature, $kT_{\rm HBB}$ is in the range of 4.5 -- 5.9 keV.
This HBB feature has been detected in some spectra of these five sources only in the IS (for the photon index $\Gamma>1.9$)  
using different X-ray telescopes. 
We also demonstrate that a timescale of the HBB-feature is of orders of magnitude  shorter  than the timescale of the iron line and its edge. That  leads us to conclude that these spectral features are formed in geometrically different parts of the source and which  are not connected to each other.  Laurent \& Titarchuk  (2018) demonstrated a presence  of a gravitational redshifted annihilation 
line emission in a BH using the Monte-Carlo simulations and therefore  the  observed HBB hump  leads us to suggest  this feature is  a gravitational redshifted annihilation  line observed   in these  black holes.}
\end{abstract}

 \keywords{accretion, accretion disks---black hole physics---stars:individual (Cyg~X--1, GX~339--4, GRS 1915+105, SS~433, V4641~Sgr):radiation mechanisms: non-thermal---physical data and processes} 
%

\section{Introduction}

High-energy astrophysics of compact objects offers a unique opportunity to take a peek at the properties of matter under the \texttt{most extreme conditions} because they produce a variety of observational manifestations which can be used to verify a theoretical model. The progress of the theoretical physics studying matter under extreme conditions is essential for creating correct models of compact objects and for adequate interpretation of available observations.

A study of the spectral variability properties of X-ray binaries is  a valuable 
source of information on the physics of accretion processes and the fundamental 
parameters of a black hole (BH). Among the observable  { signatures} 
of  a BH  was a discovery of the photon index versus quasi-periodic oscilation (QPO) frequency  
(and mass accretion rate)  correlations followed by the index saturation 
[see Shaposhnikov \& Titarchuk, 2006; 2009, hereafter ST09].
These correlations were revealed during the spectral transitions from  the low/hard state (LHS) through the intermediate state (IS) to the high soft state (HSS)  in X-ray binaries (see, for example a review by McClintock et al. 2009).

{{
In Galactic and extragalactic BHs, the X-ray spectra in the HSS are dominated by a thermal 
blackbody (BB) component that peaks in the  0.1 - 2 keV range
and in addition, these BHs  show a strong  power-law  component  
in the LHS.  Thus the energy 
spectrum is dominated by a hard Comptonization component  convolved with a weak thermal component in the LHS. The spectra of  BHs in the LHS are presumably a result of the thermal Comptonization (upscattering) of soft photons, originated 
in a relatively weak accretion disk, off electrons of the hot ambient plasma (see, e.g., Sunyaev \& Titarchuk 1980 and Titarchuk 1994, hereafter ST80 and T94, respectively). 
The IS is a transitional state of the source when it undergoes   a spectral transition  between the LHS and HSS.  The IS spectra are usually  characterized by the photon indices about 2 or higher than that.
The soft BB component of these spectra is interpreted in terms of a geometrically thin, optically thick accretion disk around the compact object 
(see Shakura \& Sunyaev 1973, hereafter SS73).  Energy dissipation within the disk should be responsible for the observed 
soft BB emission. The Comptonization continuum is believed to originate in the Compton cloud (CC) due to   upscattering 
of the thermal disk  photons by the electrons of a hot corona or a hot ambient plasma (ST80; Haardt 1993; Dove et al. 1997; Dauser et al. 2013). 
In addition to the LHS, IS and HSS sometimes very soft state (VSS) is observed in which the BB component is dominated in the source spectrum and the power-law (PL) component is either very weak or absent at all. Possibly, in this state, the PL component is screened from the observer by the disk body, which is why the spectrum becomes so soft. The bolometric luminosity in the VSS is a factor of 2 lower than that in HSS. 

There is also  an interpretation that some of these spectra consist of  
a so called reflection component which is formed due  
to downscattereding   of the Comptonization radiation in the relatively cold  accretion disk.
The X-ray reflection effect from accretion disk or relatively  cold atmosphere of a companion in the close  binary   has been studied for more than four 
decades (see, e.g.,  Basko et al. 1974; McClintock et al. 2014; 
Ludlam et al. 2017a,b). 

The current paradigm is that the original power-law radiation illuminates the surface of the accretion 
disk. Then the X-ray photons interact with the material providing diverse atomic features. These can be caused by both  via absorption (in most cases as edges) and emission (in a form of fluorescence lines and radiative recombination  continuum). For example, the   reflection (downscattering) 
component can provide a direct information about structure, temperature, ionization 
stage, and composition of the gas in the accretion disk [see also the recent papers on this subject by Garcia et al. (2013); Tomsick et al. (2014) and Walton et al. (2016)].   Sometimes, the presence of the Fe K-shell fluorescence emission and  the absorption K-edge observed in the 6 -- 8 keV energy range can be indicative to  the reflection effect.

Furthermore, it was suggested that this  reflection  effect
occurs near a BH and the line photons suffer Doppler effects, light bending, and gravitational redshift, 
which produce a skewed line profile with a red wing that can be extended to  lower energies 
(e.g., Fabian et al. 2000; Fabian \& Vaughan 2003; Reynolds \& Nowak 2003; Dov$\check c$iak et al. 2004; Miller et al. 2008; Steiner et al. 2011; Reynolds et al. 2012; Dauser et al. 2012). 
{ However, }
Laurent \& Titarchuk (2007), hereafter LT07,  (\S 3.2 there), study in  detail an effect of 
the downscattering and analytically prove  that the characteristic  downscattering  hump is only formed { when } 
the spectral index, $\alpha$ of the incident spectrum  is essentially less than one  (or the photon index $\Gamma<2$),  see also an illustration of this effect  in ST80, Fig.~10  there. Thus all { possible detections} 
of so called the reflection (or downscatered) hump in the case  of $\alpha>1$ (or $\Gamma>2$) require  a different interpretation. 

Laurent \& Titarchuk (2018), hereafter LT18 
argued that the photon-photon interactions  occurred in the vicinity of a BH should  lead
to  the pair production there because  photon energies reach a limit of 511 keV.  
The created positrons interact with accreting electrons  and as a result the annihilation 
line photons are generated  over a relatively narrow shell which is  a few hundred meters near a BH 
horizon for a  10 solar mass BH. An Earth observer can see this annihilation line only as a broad spectral feature  at the gravitationally redshifted energies with the red shift, z about 20 since the line is formed over the  hundred meter shell near a BH horizon.
As a result the observed line energies $E_{obs}\sim511~ {\rm keV}/20\sim$ 20-40 keV.  The shape of this annihilation line feature can be approximated by a hot blackbody (HBB)  of a color temperature,  
$kT_{HBB}\sim$ 4 -- 6 keV. However, the line can be smeared out due to multiple scattering if the Klein-Nishina  optical depth of the converging flow at 511 keV is greater than 2 which unavoidably  complicates its detection.  

{
This   HBB, 
additional  to the Comptonized emission and iron line   components 
were  reported in a number of Galactic microquasars. 
Specifically, Titarchuk \& Seifina (2009; hereafter TS09),  
found that eight intermediate state (IS)  spectra of GRS~1915+105 (during the 2005--2006 flaring events) required this HBB component.
Furthermore, Seifina \& Titarchuk 
(2010; hereafter ST10) found 24  spectra of SS~433  in the IS 
during radio outburst decay which  had a strong HBB  of color temperatures    
 of 4 -- 6 keV.

A similar HBB  components have been detected from  other   BHs.
In particular,  Koljonen et al. (2013) 
analyzed  X-ray data of Cyg~X--3   obtained by $Swift$, {\it RXTE} and INTEGRAL   during 
the 2006 flaring episodes (see also Shrader et al. 2010). They found that  X-ray variability during major flaring events can be related to   the Comptonization and HBB  components. It is interesting that this variability of the HBB  component  correlates with a  change of X-ray and radio fluxes.  

Tomsick et al. (2005) reported a strong hump at $\sim$20 keV in some
{\it RXTE} spectra of 4U~1630--47, which could not be satisfactorily fitted using  their
spectral model which was consist  of the multicolor disk, cutoff power-law and Gaussian component. 
Later,  Seifina et al.  (2014) also found  a small positive excess at 20 keV  but for
only one particular {\it RXTE} observation (ID = 50135-02-03-00)  previously studied by  Tomsick et al..

Mineo et al. (2012)  revealed that X-ray spectra in the so called heartbeat  state of GRS~1915+105 can be fitted by a model which  includes the  HBB of a color temperature of 3 $-$ 6 keV. The spectra from  BH soft X-ray transients, GS~2000+25, GS~1124--68 and XTE~J1550--564 were successfully approximated by an additional thermal Comptonization component which color temperatures are in the 2 -- 4 keV range and high optical depth $\tau\sim 5$ (Zycki, Done \& Smith 2001). These findings  raise a fair question on the nature of the HBB component as an intrinsic property of BHs.
}

We present a further study  of this spectral feature using the data of  {\it RXTE}, INTEGRAL 
and {\it Beppo}SAX for  the Galactic BHs: Cyg X-1, GX 339-4, GRS~1915+105, SS~433 and V4641~Sgr during their state  transition from the LHS to the HSS through  the IS. 
We investigate   variability  of spectral properties in  short time scales, particularly  for  GRS~1915+105. 
We focus on the LHS-IS transitions.  
As we have  already aforementioned, during such a transition phase, TS09   registered a broad excess hump centered at 20 keV for the {\it RXTE} energy spectra of GRS~1915+105 and later, ST10  found this feature in the   {\it RXTE} and INTEGRAL spectra of SS~433 with  time exposures up to $\sim$3 ksec.
%
It is  interesting how these spectral 
characteristics change for a shorter  exposition. 

 As we have already pointed out, the HBB humps in X-ray spectra are well known for many BHs
 { [\cite{zycki01,Koljonen13,Mineo12}, {Tomsick et al. (2005)}, TS09, ST10].}
However, the nature  
of these humps can be different depending on physical conditions. 

In this paper  we concentrate our efforts to demonstrate   an observational manifestation of the HBB hump  
in X-ray spectra of
five BH binaries,  Cyg X-1, GX 339-4, GRS~1915+105, SS~433 and V4641~Sgr. 
In \S \ref{data} we provide details of the data analysis while in \S \ref{sp_analysis} we present description of the spectral models used for fitting of these data and provide a reader  details of  timing analysis. In \S\S \ref{obs_transition} --\ref{theory} we focuse on observational results and their interpretation. In  \S 6  we  discuss  the  main results of the paper. 
In \S \ref{conclusions} we make our final conclusions.

 \section{Observations and Data Reduction \label{data}}


{For a study presented in this paper we have analyzed archival data for five BH transient  
sources, Cyg~X--1, GX~339--4, GRS~1915+105, SS~433, and V4641~Sgr, observed with {\it RXTE}. 
We used data from the {Proportional Counter Array} (PCA) and {High-Energy X--Ray 
Timing Experiment} (HEXTE) on board of the {\it RXTE}~\citep{bradt93}. 
In Tables \ref{tab:data_cyg_x-1}--\ref{tab:data_v4641} we give 
a general information on each observation used in our analysis.
}
In the  paper we { also} used the results of our previous investigations (ST10 and TS09) 
based on the archival {\it RXTE}~data   for SS~433, {V4641~Sgr} and GRS~1915+105 
during  { 1999}, 2004, and 2005 -- 2006 outbursts, respectively. { Previously,} some of these observations 
show an additional component, required to fit the energy spectra of SS~433 and GRS~1915+105. 
 We carried out  a detailed reanalysis for these  and other BHs.   
We also {revealed} 
{Cyg~X--1, GX~339--4, and} GRS~1915+105 
in different spectral states 
and  
studied an evolution of X-ray energy spectra {of GRS~1915+105, in detail, during  the IS
over shorter time intervals.}

 Also we analyzed the archival data 
of V4641~Sgr collected by { {\it RXTE} and} {\it Beppo}SAX  (Boella et al. 1997), which were obtained 
{ in the periods of  February $-$ September, 1999 and}  March 13--14,  1999, respectively.
{Previously}, these  data were also analyzed, { e.g., by Wilms et al. (2005); Shaposhnikov and Titarchuk (2006, 2007, 2009); Tomsick et al. (2018) (for Cyg~X--1); 
Tomsick et al. (2009); Shaposhnikov and Titarchuk (2009) (for GX~339--4);} 
Rodriguez et al. (2008); TS09; Filippova et al. (2006) (for GRS 1915+105);  Miller et al. (2002), Nandi et al. (2005)  (for SS~433) and in't Zand et al. (2000) (for V4641 Sgr). 

Standard tasks of the LHEASOFT/FTOOLS 6.24 software package were utilized for data processing  using methods recommended by the {\it RXTE} Guest observer facility according to 
the {\it RXTE} Cook Book.\footnote{http://heasarc.gsfc.nasa.gov/docs/xte/recipes/cook\_book.html}
For our spectral analysis we used PCA {\it Standard 2} mode data, collected 
in the 3 $-$ 20~keV energy range. The standard dead time correction procedure 
has been applied to the data. To construct broad-band spectra, we used  the HEXTE data.
We  subtracted the background corrected  in  off-source observations. 
 Only HEXTE data  in  the 20$-$100~keV energy range were applied for our spectral analysis in order 
to exclude the channels with largest uncertainties. 
 The HEXTE data have been re-normalized based on the PCA. The data are available through the GSFC public 
archive.\footnote{http://heasarc.gsfc.nasa.gov} These observations are listed in 
Table~\ref{tab:data_cyg_x-1} (for Cyg~X--1), 
Table~\ref{tab:data_gx} (for GX~339--4),  
Table 6 of TS09, {Table~\ref{tab:data_grs} of the present paper} { (for GRS 1915+105)}, 
in Tables 4 -- 6 of ST10 (for SS~433) and Table \ref{tab:data_v4641} of the present paper (for V4641 Sgr).
 We give a special attention  to observations where this  additional excess in  the energy spectrum 
was registered (TS09, ST10).

In addition we  analyzed the INTEGRAL/IBIS/ISGRI spectra in the 2004 flaring stage  
of SS~433, coordinated with the {\it RXTE} observations. We have used the version 10.2 of the 
Offline Science Analysis (OSA) software distributed by the INTEGRAL Science Data Center 
(ISDC, http://isdc.unige.ch; see Courvoisier et al., 2003). 
For all considered objects we  used  the public 
data from the  All-Sky Monitor (ASM) 
on the board of  {\it RXTE}  \citep{bradt93} {in order to obtain  additional diagnostics of the source activity}. The ASM light curves (in the 2--12 keV energy range) were 
retrieved from the public {\it RXTE}/ASM archive\footnote{HEASARC, http://xte.mit.edu/ASM\_lc.html}.


The  monitoring {\it Ryle} Radio Telescope (15~GHz) data in the 2005 -- 2006 period were kindly provided by  Guy Pooley. 
The technical details of the radio telescope were described  by~\cite{pf97}.
We   analyzed the SS~433 spectra   from the 2004 outburst which we coordinated 
with the  simultaneous radio observations. The monitoring RATAN-600 Radio 
Telescope (2.3 GHz) data during the March 2 -- April 12, 2004 period were available through the 
public archive\footnote{http://www.sao.ru/cats/satr/XB}. 
We used three {\it Beppo}SAX detectors, Low-Energy and Medium-Energy Concentrator Spectrometer 
(LECS and MECS, respectively) and a Phoswich Detection System (PDS) detectors to 
analyze  the V4641~Sgr data. The SAXDAS data package was utilized for performing data. We 
processed the spectral analysis in the good response energy range with taking into account 
satisfactory statistics of the source: the 0.4 -- 4 keV for LECS, 1.8 -- 10 keV for MECS and 15 -- 
150 keV for PDS. 

\section{Spectral and timing analysis \label{sp_analysis}}

To fit the broad-band energy spectra of these sources we used {two different spectral models: i) 
an additive {\tt XSPEC} model consisting of  the Reflection ({\tt relxill})} component (see Garcia et al. 2014) combined with 
{\tt diskbbody} component and  a {\tt XSPEC} model 
consisting of  the Comptonization (BMC or COMPTB) component (see Titarchuk et al. 1997 and 
Farinelli \& Titarchuk 2011 correspondingly) and the iron line  {(Gaussian)} components. 
In the latter case for {all} 
 spectra we also added a {blackbody} (BB) component, for which the best-fit color temperature 
was in the range of  4.5 $-$ 5.9 keV. 
A multiplicative {\tt tbabs} 
model takes into account  absorption by neutral 
material. The {\tt tbabs} 
model parameter is an equivalent hydrogen column $N_H$. 


In Figure \ref{geometry} we present our considered  scenario, for 
 the innermost part of a BH source. We assume that  accretion onto a BH is described
by three main zones: a geometrically thin accretion disk (e.g. the standard Shakura-Sunyaev 
disk, see SS73), a transition layer (TL), which is an intermediate 
link between the accretion disk, and a converging (bulk) region (see Titarchuk \& Fiorito 2004), that is assumed  to exist, at least, below 3 Schwarzschild radii, $3R_S = 6GM_{bh}/c^2$. Furthermore, we 
take into account a relatively thin layer near a BH horizon, where the photon-photon 
interaction results in the pair production (LT18). In this case
the created positrons interact with accreting electrons there and therefore the annihilation 
line photons are possibly created and distributed over the relatively narrow shell near a BH 
horizon. An Earth observer can see this annihilation line only with the gravitationally 
redshifted energy for which  the corresponding redshift,  $z\gg 1$. Typically, 
$z\sim$20 for the pair production region. As a result,  $E_{obs}\sim E_{511}/{z}\sim$~25 keV. Formally, this  feature can be fitted   by  a  BB  XSPEC model component  of the  temperature $kT_{BB}\sim$ 4$-$6 keV. 

The spectral model parameters are the equivalent hydrogen absorption 
column density $N_H$; the spectral index $\alpha$, 
a color temperature of the HBB,
$kT_{HBB}$, normalization N$_{HBB}$ of  the HBB component,
 $\log A$ 
related to the Comptonized fraction $f$ 
[$f={A}/{(1+A)}$]; the  color temperature and normalization of the seed photon blackbody component, $kT_s$ and  N$_{bmc}$ 
{(or N$_{comp}$)}, respectively. 

We should emphasize that we use the XSPEC {\tt comptb} component (see Farinelli et al. 2007) 
instead of the {\tt bmc} when the high energy tail is clearly presented in the data. But at energies 
below this exponential tail the {\tt comptb} and {\tt bmc} are identical.  

{ 
{As one can see,  the { BMC} and {Comptb} models are 
the first principal models using  only 4 and 5 parameters, respectively, which take into account contribution of the seed BB (two best-fit parameters: the seed photon temperature and normalization), the Comptonized component 
(the spectral index, in the BMC case of  or the spectral index and cutoff energy in the COMPTB case 
and a fraction (relative normalization)  of this Comptonization component. If we trace the HBB hump in the residual, we use another two best-fit parameters of this component.  
}

{ For our Comptonization  model,} we 
found that the color (seed) photon temperature, $kT_s$ 
is about 1 keV for all 
considered data and thus we fixed values of $kT_s$ 
at 1 keV (see also TS09, ST10). An equivalent hydrogen absorption column density was fixed at the level of 
{$N_H = 6\times 10^{21}$ cm$^{-2}$ (Tomsick et al. 2018) for Cyg~X--1,  
$N_H = 5\times 10^{21}$ cm$^{-2}$ (\cite{Kong02}, Garcia et al. 2017) for GX~339--4, and }
$N_H = 5\times 10^{22}$ cm$^{-2}$ (Trudolyubov 2001), 
1.2$\times 10^{23}$ cm$^{-2}$ (Filippova et al. 2006) and 2.3$\times 10^{21}$ cm$^{-2}$ (Dickey \& Lockman 1990) for 
GRS~1915+105, SS~433 and V4641~Sgr, respectively. 
When the parameter $\log(A)\gg1$ we fix  $\log(A)=2$   
because the Comptonized illumination fraction $f=A/(1+A)\to$1 and variation of $A$  does not improve the fit quality any more.

Similarly to the ordinary bbody XSPEC model, the  BMC and Comptb normalizations    
is a ratio of the source (disk) luminosity L to the square of the distance $d$: 

\begin{equation}
N=\biggl(\frac{L}{10^{39}\mathrm{erg/s}}\biggr)\biggl(\frac{10\,\mathrm{kpc}}{d}\biggr)^2.
\label{bmc_norm}
\end{equation}  
In Tables 5-13, 15-18 
we demonstrated fit qualities of  our  spectral model. 
The {\it RXTE} light curves were analyzed using the powspec task from FTOOLS 6.24. The 
timing analysis of  {\it RXTE}/PCA data was performed in the 15 $-$ 30 keV energy range using the event 
mode for the 91701-01-49-00 observation of GRS 1915+105, { where we found a noticable  excess 
at 20 -- 40 keV in terms of  our model}. The time resolution for this observation is 1.5$\times 10^{-5}$ s. 
We generated the power density spectra (PDS) in 0.1 -- 500 Hz frequency 
range with 0.01 second time resolution. We subtracted a contribution due to Poissonian 
statistics and Very Large Event Window. To model PDS we used { QDP}/PLT plotting package. 

The PDS continuum shape for  BHs in the LHS and  IS has  usually a band-limited noise 
shape~\citep{tsa07}, which is well presented by an KING model in QDP/PLT. To fit QPO features we 
use Lorentzian shape. We quote the Lorentzian centroid as a QPO frequency.

\section{Results  \label{obs_transition}}

{
Below we present,   in details,  results for all studied sources using different spectral models. 
} We compared the results using the Comptonization and reflection 
models and  found a presence of an additional component (above  20 keV) in the
spectra for  all sources  using the  our Comptonization model. 
It is interesting that additional excess of radiation of 20 keV does not arise when we use the reflection model.
Moreover, the Comptonization model  better
describes the spectral evolution of considered BHs than the reflection model. Thus below  we
systematically elaborated  this spectral study.

We analyzed the data sets for five sources: Cyg~X--1, GX~339--4, GRS~1915+105, SS433 and
V4641~Sgr. Among them, for Cyg X--1, GX~339--4 we have much more extensive  data  than those  for others. 
 Therefore, we use these two sources for a more detailed study of the  spectral  transitions from the LHS to IS and from the IS to HSS.   It is remarkable that  three other sources (GRS~1915+105, SS433, V4641~Sgr) fit well the resulting picture of the spectral behavior.

\subsection{Cyg~X--1 \label{res_cyg_x-1}}

{
Cyg~X--1 is a classical BH source  and one of the sources most extensively observed by {\it RXTE} (Wilms et al., 2005; ST06, ST07, ST09). Recently, 
a broadband spectra of Cyg~X--1 applying  $Nu$STAR and $Suzaku$ make it possible to improve our knowledge on the iron line spectral region of this source. 
Previously, these spectra of Cyg~X--1 were also analyzed by Tomsick et el. (2014, 2018), 
Walton et al (2016) and Basak et al (2017) using so-called 
{reflection} models. They claimed    their model  is in  a good agreement  with for all spectra observed throughout 
spectral states of the source. 

}

We   investigate   the Cyg X-1 spectra using  our Comptonization model. {
 We analyzed 832 {\it RXTE} observations of Cyg~X--1 made between February 1996 and July 2005 using the public archieve 
(see Table~\ref{tab:data_cyg_x-1}). Using our Comptonization model we  found that the source was in   four spectral states: the LHS, the IS,  the HSS and the VHS. 
We revealed  a significant 
positive excess (so-called  HBB hump, at 15--40 keV) in some Cyg~X--1 spectra when we fitted the  data applying  the model 
{\tt tbabs*(comptb+gauss)*smedge}. The shape of this additional component can be 
described  by a simple {\tt bbody} (HBB) model with a color  temperature of  $\sim 4 - 5$ keV.
As a result we use the model {\tt tbabs*(comptb+gauss+bbody)*smedge} and
apply it for all spectra of Cyg~X--1 in order to describe the source evolution. 
We studied how the spectral parameters change during the state transition  and also find  significant HBB normalization,   (see e.g. Table 5)
 detected mostly in the IS and sometimes in a so-called LHS--IS spectral transition.  Cyg X-1  demonstrates  a lot of spectra of this type.
Furthermore, we revealed a significant HBB component  even for $\Gamma < 2$. But what is remarkable that the hump was observed in the range of  $1.8<\Gamma < 2$ and only in those cases when the electron temperature was lower ($\sim$ 20 keV, see Table~\ref{tab:fit_50119_cyg_x-1} at MJD=51908 (50109-01-04-00).
However,  for the same index range  
($1.8<\Gamma < 2$), but for higher plasma temperature ($kT_e\sim 50$ keV), the HBB  was very weak  for $\Gamma \sim 1.8$ 
[see 50109-03-13-00 (MJD=51919)].

We also detected a hump for the $\Gamma =$2--2.5. 
The hump becomes stronger when $\Gamma$ increases from 2 to 2.3 and then it is getting weaker  when $\Gamma$ reaches 2.8. 
At high $\Gamma$, the detection of the HBB  falls below the detection limit.  For such a case  we fix  normalization  of the HBB
at $N_{HBB}=0.2$ L$_{37}$ erg/s/cm$^2$. 
For Cyg~X--1, the necessary observational conditions for detection of the HBB  are  the 
$1.8 <\Gamma <2.5$ range   and relatively  low electron temperatures ($kT_e$).
We should notice  that Cyg~X-1 is characterized by different levels of X-ray  luminosity  at different 
epoches and also different index values (ST09). When the   luminosity level is higher, 
the source can achieve higher levels of the index, and vice versa (see Shaposhnikov \& Titarchuk, 2006 hereafer  ST06).  In Figure \ref{four_sp_cyg}  using our Comptonization model we illustrate how Cyg X-1 evolves from the LHS (black  line and observational points) to the VHS (blue line and points) through the IS (red line and points) and the HSS  (green line and points), see also Table \ref{tab:fit_10512_cyg_x-1}-\ref{tab:fit_60091_cyg_x-1} . 

The shape of this additional component can be 
described  by the high temperature  blackbody (HBB) component (Fig. \ref{sp_compar_cyg_refl_bbody}, see upper  panels). As it is seen from this Figure the maximum of the HBB component is observed when the source is in the IS.
But this HBB component cannot  be determined using the reflection model (see lower panels). 

It is important to emphasize that the HBB hump has its centroid energy around 20 keV 
similar to that found in the Monte-Carlo simulations of the redshifted annihilation line of 511 keV (see LT18).
Specifically, we detected 169  events (among 832 ones) in which the HBB   was clearly seen 
during the IS (see details in Tables \ref{tab:data_cyg_x-1}, \ref{tab:fit_10512_cyg_x-1}--\ref{tab:fit_60091_cyg_x-1}). 

In Figure \ref{index_vs_soft_flux_comptb_refl_mod0}  we demonstrate  the  photon index $\Gamma$ vs the {\it RXTE} soft X-ray flux (3--10 keV) correlation for Cyg~X--1 using the Comptb model (left panel) and  the reflection model (right panel). One can clearly see in the left panel that $\Gamma$ evolves from values of  1.5 (in the LHS) to those of 2.2 (in the  IS) with a clear signature of  the index saturation at 2.2. It is worth noting another plateau  at  about 2.6, seen in this Figure,   was found by Shaposhnikov \& Titarchuk (2006), hereafter ST06. In fact, the index plateau  value depends on the plasma temperature $kT_e$  of the converging flow 
(see details in  LT99 and LT11).

On the other hand one cannot see any variation of $\Gamma$ with the soft X-ray flux using the reflection model  (see the right panel).   The next natural question  arises what typical time scales for the iron line-edge and the HBB are and how these features  are related between each other.   
In Figure \ref{time_scale_for_hbb_and_gaussian}  we clearly demonstrate that the time scale of the line features is of two  orders of magnitude larger of those for the HBB.  These observational facts definitely indicate to the different   origins of these spectral components (see more discussion in \S  5). 
 
\subsection{GX~339--4 \label{res_gx339-4}} 

{
GX~334--9 shows a similar X--ray behavior to Cyg~X--1. The source  spends most of the time in the LHS  like a faint persistent source. It occasionally shows  an outburst  with  a transition from the  LHS to the HSS. Usually, during the HSS, 
it becomes brighter by a factor of 5--100 in the 2--10~keV band. In Figure \ref{five_sp_gx} we show a spectral  evolution 
of GX 334--9 from the LHS to the the very high state (VSS).
The detailed analysis  of the X-ray spectra in GX~339--4  is essential to identify 
different spectral states in this source.   In addition to the LHS and the HSS, GX~339--4  also exhibits the  softer spectrum  and higher luminosity in VSS  than that in the HSS. 

Similar to the analysis of Cyg X-1 (see \S~\ref{res_cyg_x-1}), we also applied the {\tt relxill} model to the 
{\it RXTE} spectra of 
GX~339--4.  We present the spectral evolution of GX~339--4 using this reflection model in Figure \ref{sp_compar_gx_refl_bbody2}.    Best-fit results are also shown in Table~\ref{tab:fit_gx339_reflection}. 
Applying  the {\tt relxill} model we found  the very high iron abundance of nearly 3 and 10 times of the solar abundances (see  the 92035-01-03-00 and 70109-01-07-00 observations).  
Although some other X-ray binaries may have supersolar abundances, it is difficult to explain why $A_{Fe}$ would be variable for the same source. It is also strange that   the derived binary inclination, $i$  varies from 20$^{\circ}$ to 90$^{\circ}$. 

We  continue  to study  the  GX~339--4 spectra and analyzed them  using  our Comptonization model.
We use 135 {\it RXTE} observations  made between  February 2001 and June 2007  presented in  the public archive (Table~\ref{tab:data_gx}).  The best-fit results are shown  in Tables~\ref{tab:fit_70109_gx} -- \ref{tab:fit_92428_gx}. We found 27 spectra with a significant HBB  (among 135 spectra)   during the IS as well as in the LHS--IS  transition (see details in Tables \ref{tab:data_gx}, \ref{tab:fit_70109_gx}--\ref{tab:fit_92428_gx}).  It is interesting  how the HBB evolves when GX~339--4 evolves from the LHS to the IS and vice versa (see upper panel in Fig. \ref{sp_compar_gx_refl_bbody2}). 

We also  demonstrate this evolution versus $\Gamma$ and $kT_e$ in Figure ~\ref{sp_evol_gx_LHS-IS}. Data are taken from  the {\it RXTE} observations 70110-01-97-00 ($kT_e=50$ keV, $\Gamma$=1.7,  LHS, left), 
90418-01-01-01 ($kT_e=27$ keV, $\Gamma$=1.7, the LHS, left center), 60705-01-69-01 ($kT_e=25$ keV, $\Gamma$=1.97, 
LHS$\to$IS, right center), and 70110-01-08-00 ($kT_e=18$ keV, $\Gamma$=2.0, LHS$\to$IS, right). 
Using {yellow} shaded areas we indicate the HBB component which presence or absence 
can be related to the index value, $\Gamma$  and consequently to a value of  $kT_e$. 

When GX~339--4 is in the LHS ($\Gamma\sim$ 1.7, two left panels in Fig. \ref{sp_evol_gx_LHS-IS}) one does not see the HBB  for $\Gamma=1.7$ for 
relatively high electron temperatures ($kT_e=50$ keV and $kT_e=27$ keV).
However, 
when this source is in the  LHS--IS transition ($\Gamma\sim$ 2), we see a significant increase
of the HBB which is particularly  strong when the electron temperature is lower ($kT_e=18$ keV, right panel) than that for the higher one ($kT_e=25$ keV).  We found that the HBB  is mostly detected in GX~339--4 spectra  when $\Gamma$ changes from 1.8 to 2.5. 
 For example, in the LHS-IS transition for 
$\Gamma\sim 1.85$ with a lower electron temperature ($kT_e=23\pm 8$ keV, id=70109-01-06-00, MJD=52400.83, 
see Table~\ref{tab:fit_70109_gx}) we detect a stronger HBB component ($N_{HBB}=4.5$ L$_{37}$ erg/s/cm$^2$) than that 
for  higher electron temperature ($kT_e=37\pm 2$ keV, id=60705-01-56-00, MJD=52710.71 and lower normalization, 
$N_{HBB}=0.2$ L$_{37}$ erg/s/cm$^2$,
 see Table~\ref{tab:fit_70128_gx-1}).

In Figure \ref{index_vs_soft_flux_comptb_refl_mod3}  we show the photon index $\Gamma$ evolution versus the 3-10 keV flux  using the Comptb  model.
As one can see $\Gamma$ increases and then it saturates when the flux   increases using the Comptb model .


}

\subsection{GRS~1915+105 \label{res_grs1915}}

{ Previously, we revealed the presence  of the HBB  spectral component 
 in several broad-band spectra of GRS~1915+105} (1996 -- 2006) analyzed by TS09 where we found the HBB hump component around 20 keV in eight IS and HSS spectra among 107 available spectra of GRS~1915+105.}
{ In the present paper we reanalyzed { these data of eight observations for  GRS~1915+105} 
(see Table~\ref{tab:data_grs}) 
to check out our previous {conclusions} 
on detection or nondetection the HBB feature. It is worth noting that }
the HBB  feature was { well fitted}  by a $\sim$4.5 keV  BB  profile  
(see Fig. 4 and Table 7 in TS09).

 
{


 In Figure \ref{three_sp_grs} we present the best-fit spectra for four different spectral  states observed in GRS 1915+105. As it is seen there  luminosity increases when the source goes to the softer state. 
{The best-fit parameters of the spectra  
are shown in Table~\ref{tab:fit_grs}. 
}
In Figure~\ref{gam_2005} { (top panels)} we demonstrate the photon index, $\Gamma$ plotted versus Comptb normalization 
(left) and hardness-flux diagram (right) for the same, 2005 -- 2006 transition of GRS~1915+105 
analyzed  using tbabs*(comptb+gauss+bbody)*smedge model.
{ From these plots we can conclude that }
$\Gamma$ monotonically increases with the normalization parameter (proportional 
to the  mass accretion rate, $\dot M$) when the source makes  a transition from the hard to soft state and the soft X-ray flux (or $\dot M$) grows. 
Furthermore, $\Gamma$ demonstrates the saturation at  $\Gamma_{sat}\sim 3$.  
It is evident  that our Comptonization model fits the source spectra for all spectral states 
with good statistical significance (see  right column of Table~\ref{tab:fit_grs}). 
We should emphasize  that the adopted spectral model shows a very good performance
for all cases used in our analysis. 
We should note that  the data related to the IS eight observations  are fitted 
 using a model, tbabs*(comptb+gauss+bbody)*smedge  which is  especially
needed to decribe all residuals of the data (8 cases among 32 spectra in
the present paper, and 8 cases among 107 spectra in TS09). Applying  this  model we found  a characteristic hump around 20 keV (see the central bottom panel of 
Fig. \ref{sp_compar_grs_refl_bbody}) which can
be fitted by a BB shape of a color temperature around 4.5 keV (HBB)
(see Table \ref{tab:fit_grs} for values of the best-fit parameters). This  
HBB component is strong in each of these  observations, and its
EW varies from 300 to 700 eV (see also Table  {\ref{tab:par_bbody}).

Figure \ref{lc_grs} shows the 2005 -- 2006 evolutions of the flux density $S_{15GHz}$ at 15 GHz 
({\it Ryle} Telescope, top), {\it RXTE}/ASM count rate ({second from the top}), the Comptb normalization, and 
 $\Gamma$ (see two bottom panels, respectively) during the 2005 -- 2006 middle and  decay 
transitions of GRS~1915+105 (MJD 53690 -- 53855).
{ {Green} and {red}} points (in two last panels) correspond to normalization and  $\Gamma$.
All vertical strips  mark the time intervals (up to $\sim$3 ksec), in which the HBB 
component was detected. For these observations, it is particularly interesting 
how spectral characteristics changed in the case of  shorter time exposition. 

Now  we try to understand using as an  example of GRS~1915+105 
how small a lifetime of the found HBB spectral feature  and thus estimate its lower limit.
 In Tables {\ref{tab:par_bbody}$-$\ref{tab:fit_grs_laor} we  show  a behavior of spectral parameters in  the  10$-$30  s intervals   (see  also pink vertical strip,  related to the 91701-01-49-00 observation,   
 MJD=53771.51$-$53\-772.1  in Fig. \ref{lc_grs}). 
 GRS~1915+105 is a bright  X-ray 
source   (not like   much fainter sources SS~433 and V4641~Sgr, see below), allows 
to study the short time exposition spectra with a significant S/N ratio.

The observations  were carried out 
for  the 15$-$30 s  {intervals} 
interrupted by dead time { periods}. 
For { PCA and HEXTE} 
detectors, so called start and stop times of these short observational intervals can be different. {Therefore, we applied} 
the HEXTE time interval distribution in order  to select only simultaneous { (for both PCA and HEXTE)} observational intervals    
and { } synchronize them {to make  a consistent   spectral analysis.} 
We found that  the PCA spectra are { well} matched with 
the HEXTE spectra for the 91701-01-49-00 observation { where the HBB hump  is clearly detected}. The corresponding 
intervals obtained for the broad-band PCA\&HEXTE spectra are listed in Table \ref{tab:par_bbody}. 
The HBB feature has been found { only} in the first eleven  {spectral bins} 
of GRS 1915+105 among all available time bin intervals.
For  these  time segments the spectra { are} 
 well fitted using of the HBB component.  We presented 
the best-fit parameters  for these  HBB time segments of the GRS 1915+105 observations { in } 
Tables \ref{tab:fit_grs}$-$\ref{tab:fit_grs_laor}. 

{ We found that the} 
color temperature of the detected HBB feature  
 is about 4.5 keV and  equivalent width (EW) of the HBB component is in the range of 300$-$700 
eV { for these} 
first eleven {sub-segments (about 600 s)} 
of the {91701-01-49-00} 
observation. { As a result, } this HBB hump is clearly detected in the  spectrum of GRS~1915+105 
{ during first 600 s, 
which can be specified as the life time of this spectral feature.} 

{
}



{ It is interesting that } 
the observed QPO feature peaks at $\sim$1 Hz frequency (see details in TS09).
The $\nu_{QPO}$  presumably points out  { the presence of} 
the bounded region 
[the transition layer (TL)] for which this QPO frequency is an intrinsic (eigen) frequency of the TL 
and the power of the source, around 1 Hz, excites this 1 Hz QPO. 
 In fact,  $\nu_{QPO}\sim V_p/L_{tl}$ where $V_p$ and $L_{tl}$ are the proton velocity and the TL size, respectively (see Titarchuk et al. 1998). A characteristic value of  $V_p$ is about $3\times10^7$ cm s$^{-1}$ and  $L_{tl}\sim 3\times10^7$ cm and thus one can obtain that 
$\nu_{QPO}\sim1$ Hz.
  
This QPO frequency   definitely   indicates to the compactness of the TL. 
{Furthermore,} GRS 1915+105 demonstrates a change of  $\nu_{QPO}$  in the PDS during the 
LHS$-$IS$-$HSS outburst transition in the range 0.1 -- 10 Hz (Trudolyubov 2001).  These  lower frequency QPOs 
{ are usually   observed in the IS when radio flux from GRS~1915+105 decreases to 50 mJy}
(see Fig. \ref{lc_grs} here and Fig. 15 in TS09). 

\subsection{SS 433 (V1343 Aql) \label{res_SS}}
Although  SS~433 is located close to the ``disk edge-on'' position relative to
the Earth observer ($i \sim 78^{\circ}$) and it is difficult to analyze, especially in
optics, nevertheless it is a source of heated debate what  it is  in its center -- BH or NS. It is also  important to study this source in the X-ray range, whose photons
undergo less absorption. In addition, the object, most of its time, is in IS, which is
most interesting for the purposes of our work. The fact of detecting the HBB features
in the spectrum of SS433 would be a further argument in favor of a BH in this object.
{ The  X-ray spectra of SS~433 were usually  fitted using the XSPEC  Bremsstrahlung  model for which the best-fit 
temperature about 20 keV (Kawai et al. 1989; Cherepashchuk et al. 2005).}
However, 
we fitted  the  spectra of SS~433  using an additive Comptonization XSPEC model  
with a small 
modification of the emission line { components} 
 using two narrow and 
wide Gaussian line components. 
{ Moreover,  the data  require  an additional component, HBB to describe several 
spectra:   {\tt (comptb+Gaussian1+Gaussian2+bbody)*smedge}. The presence of 
the HBB
feature has been found in 24 {\it RXTE} spectra among 117 available spectra of SS~433 
(1998 -- 2006),  
listed in Tables 4$-$6 in ST10. 
{This feature has been  also found in the broad-band source spectrum combined using two simultaneous observations 
of {\it RXTE} (ID 90401-01-01-01) and INTEGRAL (rev.~336).}
We should emphasize  that  the HBB   was detected during a flare of the source.

Figure \ref{lc_ss}  illustrates ({from top to bottom} evolutions of the flux density $S_{2.3GHz}$ at 2.3 GHz (RATAN--600), {\it RXTE}/ASM count rate, the {\tt comptb} normalization, and 
$\Gamma$ during the March$-$April, 2004 period for the { rise--peak--decay} 
transition of SS~433 (MJD  53060$-$53114). Vertical strips mark the observational intervals, for which an additional HBB 
was required. The pink vertical
 strip indicates the time interval of INTERGAL which was  simultaneous with the {\it RXTE} observation of SS~433  (ID 90401-01-01-01).  The best-fit parameters for PCA/{\it RXTE}+ISGRI/IBIS/INTEGRAL spectrum are presented in  Table \ref{tab:fit_int}. 
The best-fit spectra of SS~433 shown by Seifina \& Titarchuk (2010) 
correspond to the combined 
{\it RXTE}+INTEGRAL spectrum 
and PCA/HEXTE/{\it RXTE} spectrum.
 We show the photon index vs.  normalization (proportional to the mass accretion rate) 
correlation in Fig. \ref{gam_2005} (left bottom panel).
{Using this Figure one can clearly see how $\Gamma$ monotonically increases from 1.8 to 2.3 versus   the mass accretion   rate, $\dot M$  and finally saturates at high $\dot M$ on the level of $\Gamma\sim 2.3$.
It is worth noting that in this plot we include an additional point obtained using the {\it Beppo}SAX observations of SS 433 (see \S \ref{data}). 
 
 SS~433 is a unique BH which mostly shows 
 the IS  and sometimes seen in  short time transitions 
from the IS to LHS and vice versa. A nature of the source is highly debated.
The key moment is a mass of the compact object (ranged from 1.4 to 4.2 $M_{\odot}$, Gaise et al., 2004), 
which favors to either a  BH or a NS cases. However, X-ray spectral signature of a BH as a correlation of the spectral (photon) index vs mass accretion rate developed by ST09 points to   the BH nature of the central object in SS~433 (see also ST10).

\subsection{V4641 Sgr (SAX J1819.3--2525)  \label{res_V4641}} 

Outburst events of V4641 Sgr in September 1999 are consistent  with a number of well 
separated flares of different amplidudes depending on    energy. 
The Wide Field Cameras (WFC) of 
the { \it Beppo}SAX at time MJD 51431   observed an  outburst. This outburst was examined by in't Zand et al. 
(2000) and they found  the best-fit using  the XSPEC model:  phabs(compTT + bbody + Gaussian).
 The BB temperature, kT$_{BB}$ was 2.4$\pm$0.2 keV. However, Figure 7 of  their paper 
demonstrates that the fit quality 
is not acceptable, mainly because of residuals around 20 keV. Moreover, the obtained color temperature 
of this BB component of $\sim$2.4 keV is too high for a typical   accretion disk temperature. 


Therefore, we { reanalyzed these  X-ray spectral data}. 
To do this new analysis we investigated the  {\it Beppo}SAX NFI observation (ID=20549006) 
carried out on  March 13--14, 1999. As we have already mentioned above, in a previous study of this spectrum, in't 
Zand et al. (2000) fitted the Fe K$_{\alpha}$ line region with two zero-width Gaussians (at 6.68 keV 
and 6.97 keV for Fe XXV and Fe XXVI), respectively with an unsuffcient fit quality. Miller 
at al. (2002) significantly improved the fit statistic  using the Laor line model to describe 
the Fe K$_{\alpha}$ line region. However, all previous examinations considered the 0.4 -- 10 keV energy range  only.  
In the present paper we investigated the { broader} band (0.3 -- 150 keV) spectrum of 
V4641 Sgr, adopting the Comptb  model for the continuum and the Laor line 
model for Fe K$_{\alpha}$ line region. 



We   used the {\it RXTE} data for V4641 Sgr (see Table \ref{tab:fit_4641_rxte}, first column for obs. IDs )
and analyzed  the data    
similarly to that for GRS 1915+105 and SS~433 (see upper panels in Figure \ref{sp_all_bbody}).
However, the  {\it RXTE} detectors cannot provide high quality spectra for energies below 3 keV 
while a broad energy { range} 
of {\it Beppo}SAX  allows us to reliably  determine  the Comptb parameters 
using soft energies. Due to a good spectral resolution of {\it Beppo}SAX NFI detectors one can find 
the sharp  K-edge at 7 keV in the spectrum of V4641 Sgr. In addition, we 
revealed the HBB  component with a color temperature of about 4.5 keV. 
The best-fit parameter values  are shown in Tables \ref{tab:fit_4641}$-$\ref{tab:fit_4641_rxte} 
 {while in  Fig. \ref{sp_all_bbody} 
we show the corresponding best-fit  spectra 
of V4641 Sgr during the IS  event in E*F(E) units. 
The data are  presented by 
 crosses using the spectral model:  {\tt tbabs*(comptb+laor+bbody)*smedge}. 
 The spectral model  components   are shown 
for the Comptb, Laor, and BB components (see details  of the fits in Tables \ref{tab:fit_4641}$-$\ref{tab:fit_4641_rxte}). 
 


The   determination of  the HBB hump 
can be affected by the presence of nearby Fe-line K-complex (6.5 -- 7 keV)  associated with  K-edge at $\sim$7 keV.  However,  the photoelectric absorption cross section 
sharply decreases with a photon energy [approximately $\sigma_{ph}\sim(7.8~{\rm keV}/E)^3\sigma_T$], and therefore  K-edge cannot change  a  central part (10 -- 30 keV) of  the HBB hump. This is clearly seen   
in Fig. \ref{sp_all_bbody} (lower panel) where  the {\it BeppoSAX} NFI spectrum (ID=205490061) of V4641 Sgr obtained with a high energy resolution was fitted by our Comptonization model. 

As we have already mentioned  above this V4641~Sgr  spectrum  was  investigated in detail in the previous 
studies by \cite{zand2000} and \cite{Miller02}.  But their studies were based on the soft spectrum (0.4 -- 10 keV) and therefore
{ their results differ  }
in  values of the photon index $\Gamma$. For example, In't Zand et al.  claimed that  they found  the high soft state spectrum 
{ of V4641~Sgr} dominated by the  disk component  with a weak power law of $\Gamma\sim 3$.
They also found  the absence of narrow or broad-band features in the PDS 
of MECS light curve.  On the other hand, {Miller et al.  argued about 
 an  IS  identification,  based on the obtained value of the photon index, $\Gamma=$1.3$\pm$0.1 and intermediate X-ray luminosity. 
Because of the low value of  $\Gamma$, Miller et al.  suggested that  V4641~Sgr was observed  in a {unique substate} 
of  the IS, which may differ from  similar states in other Galactic BHs. However, we studied a broad-band spectrum (0.3$-$150 keV) of V4641~Sgr and found  
a value of the photon index $\Gamma=$2.5$\pm$0.1, { which is in agreement with  } 
a typical IS { identification} (see Table \ref{tab:fit_4641}). 
{ In the radio observations of V4641~Sgr a double-sided jet structure has been  detected in VLA 
image taken on 16 September, 1999 (Hjellming et al. 1999). These events }
could indicate to a radio active state  like to GRS 1915+105 and SS~433 during X-ray outbursts. 

Now,  we  can summarize the main spectral parameters for all five objects that  
 there is   a  common condition for the detection  the HBB feature.  All these spectra where we revealed  these HBB components, are characterized by the photon index of  the Comptonized  component, $\Gamma\gax$ 2 (compare with LT18).  It is also important that we have found  using the {\it RXTE}  data for Cyg X-1 that  the time scale of the iron line features  is much longer (of order of $10^5$ s) than that for the HBB component (of order of $10^3$ s) and  which indicate that these two components are unrelated and formed in  different parts of the source. 

\section{Interpretation and discussion of observational results \label{theory}}

Before to proceed with an interpretation of the analyzed observations let us briefly   summarize 
them as follows:

i. We detected  the HBB hump in five black holes in their intermediate state (IS) and the LHS-IS, IS-HSS transitions.  
ii. The HBB hump was found in the IS spectra  and related to 
X-ray flux  of order of $10^{37}$ erg/s. 
iii. A strong iron-line complex is seen along with  the HBB broad feature and they are  unrelated in terms of time scales.
iv. Usually,  the   IS spectra are
accompanied by a moderate  radio emission. 
v. Our   model in which the continuum is characterized by only 5 parameters and therefore has  high significance. 
vi. The meaning of the  normalization  in our  model is related to the disk seed photon flux and it is proportional to the disk mass accretion rate.

It is interesting that the HBB  hump was detected  previously, for example, in the Cyg~X--1 spectra using {\it Suzaku} + NuSTAR data (e.g., 
see Fig.~2 in Walton et al. 2006; Figs. 4 and 5 in Tomsick et al. 2018). However, this hump was interpreted 
as a reflection bump  using   the {\tt diskbb + reflionx} model.  But it is well known that 
the reflection (downscattering) is uneffective when $\Gamma > 2$ (or around 2) because  of  lack of high energy photons. Thus the reflection  hump is not formed  in these cases
(see LT07).

 The loss 
of the HBB hump detection  is  usually caused by averaging  spectra of different spectral states (see ST09). Moreover, 
this feature shortly lives and extremely transient. The nature of the HBB component (or so called a reflection hump in the literature)  is still hot disputable.  So, it is reasonable to consider possible interpretations and thoroughly study the observational manifestations of this feature along outburst rise/decay in  detail.

\subsection{Observations of 
the HBB  component \label{why intermediate}}
{
  Using spectral analysis of BHs: Cyg~X--1, GX~339--4, GRS~1915+105, SS~433, V4641~Sgr observed by 
{\it RXTE}, INTEGRAL and 
{\it Beppo}SAX we found that HBB component 
 is well detected in the IS spectra. Also  this HBB   
is never detected in the {LHS} 
spectra. We find some detections of the HBB  in the LHS--IS and the IS--HSS spectra when $\Gamma$ 
changes from 1.8 to 2.8 and  $kT_e$ is around 20 keV or less 
(see e.g.   Cyg~X--1 and GX~339--4 cases).
 We also detected this  HBB  of the color temperature of 4.5--5 keV  
in the energy spectrum of GRS~1915+105 (see Figs. \ref{sp_compar_grs_refl_bbody}-\ref{lc_grs}  ).

{

However, it is important to know  if this HBB feature presents through the whole observation or not. 
We divided the  particular PCA and HEXTE simultaneous observation into subintervals and find that
the lifetime of this feature is about of 11 min in  the case of GRS~1915+105 (see also Table \ref{tab:par_bbody}). But it was absent for  following observations (12 -- 32, 
 Table \ref{tab:par_bbody}). 
 Although  the total duration of this { observation} 
 is around 30 minutes but the HBB was seen only during the first 10 minutes.


The HBB feature is well detected in the energy spectrum, when  a BH:

$\bullet$ 
 is in the active phase, namely, in the IS or during the IS-LHS and  IS--HSS transitions;  

$\bullet$ 
is characterized by a value of  dimensionless  mass accretion rate, $\dot m\sim 1$ (see LT18);

$\bullet$ 
is accompanied by moderate radio-wave emission (see Figures \ref{lc_grs}$-$\ref{lc_ss});


$\bullet$ 
is characterized by a photon index in the range of  1.8$<\Gamma<2.8$ (see 
Fig. \ref{EW_new+SS433}); 

$\bullet$ 
is observed when   X-ray luminosity in the range of $5\times 10^{36}$ --  $5\times 10^{37}$ erg/s; 


On the other hand   we associate  unfavorable conditions with the LHS and HSS, when  the HBB  is usually  absent or smeared out  (see,  as an example,  Fig \ref{sp_compar_grs_refl_bbody})
We hope that the current and future observations can easily detect this HBB  feature based on our empirical criteria.  {\bf There are many examples of the detection of such a hump-like feature in the X-ray source spectra by other authors. For example,} Walton et al.  (2016)   [Fig. 9 there] clearly demonstrate the presence of the hump in the energy band from 10 to 40 keV in the Cyg X-1 spectrum using $Nu$star and $Suzaku$ data.
Another example of the observation of the high/soft spectrum of Cyg X-1
is shown by Tomsick et al (2014). The authors also find the high/soft
spectra (see Fig. 5, panel a there) where one can see a hump
between 10 and 40 keV. 
According to LT07 
(see also ST80)  the downscattering (reflection) hump should be clearly observed 
in the hard state demonstrated by Garcia et al. 
We emphasize that the strongest effect of the down-scattering should be observed  in  the LHS. 
But if  the down-scattering hump is not formed in the LHS
 then it should  not be  formed in the reflection case too. In other words 
 in the reflection case a fraction of the reflected (down-scattered) photons, 25 \%  is much less than
 that  for the transmission case when the fraction of the down-scattered photons  is  100\%.

{
\subsection{Similarity between the upscattering photon spectra and the distribution of capital gain due to investement \label{capital}}
Titarchuk et al. (2009), hereafter TLS09, strongly argue  that there is a full similarity between the formation of the Comptonization (upscattering) photon spectrum and capital gain due to investment
[see also Dragulescu \& Yakovenko  (2001)]. Furthermore, the photon upscatering is analogous  to the Fermi particle acceleration.  
TLS09 derive the Comptonization spectrum  as a broken power-law (see Eq. A5, there) for which the spectral index of the hard tail $\alpha$  is 
\begin{equation}
\alpha = \frac{\ln(1/p)}{\ln(1+\eta)}
\label{pl_index}
\end{equation} 
where $p$ is average probability of photon scattering in the CC, or in the case of capital gain is average probability of success due to investment and $\eta$ is related to the average photon energy (money) change per scattering (investment)  
$\Delta E = \eta E$ (where $\eta > 0$ for the upscattering case or successful investment).
TLS09 demonstrated that $\alpha\approx(\eta N_{sc})^{-1}=Y^{-1}$ where $Y$ is the Comptonization parameter [see Rybicki \& Lightman   (1979)].  

Thus one can  suggest the bounded transition layer [Compton cloud (CC)]  near the central  object, a NS or a BH where the well-known Comptonization spectrum is produced (see e.g. ST80). Titarchuk \& Seifina  (2019) argue this bounded configuration (CC) is formed due to an adjustment of the Keplerian disk motion to the inner sub-Keplerian rotation of the central object. The observational change of the photon index due to the mass accretion rate  is well established (see  Fig. \ref{gam_2005}  and  ST09).   But the relative constancy of the index in the framework of the reflection  model 
(see Fig. \ref{index_vs_soft_flux_comptb_refl_mod0}, right panel)  can be only explained by unnecessary surplus of free parameters using this model. Moreover, applying the reflection model we cannot see a well-known the spectral transition in BHs (see e.g. Park et al. 2004 and ST09).  } 


{
The main portion of the X-ray 
spectrum can be described and explained by a Comptonization of the soft (disk) photons, parametrized by a photon index $\Gamma$. In its turn,  $\Gamma$ inversely proportional to the Comptonization parameter Y   is a product of upscaterring efficiency at single scattering $\eta$ and number of scattering $N_{sc}$ (see 
Eq.  \ref{pl_index} and  text below of it).  
Thus the spectral index evolution  indicates to a different  spectral state,  and   is  essential to 
constrain spectral models (see also McCintock 2005; Remilard \& McClintok 2006).

In other words  if the spectral model does not reproduce the  spectral evolution (in terms of its parameters) 
and by time evolution (LF-QPO frequency vs soft X-ray luminosity or mass accretion rate), then something is  
wrong with  this particular model 

In addition, the Comptonization model (Comptb, Comptt, BMC) is based on a transition-layer paradigm 
(TLM98), in which the LF-QPO self-consistently reproduces their frequency, and the correlation 
$\Gamma$ versus LF-QPO frequency during transitions between diffent spectral states (e.g., see Vignarca et al., 2005; Shaposhnikov \& Titarchuk, 2009; Stiele et al. 2010).

It turns out that this Comptonization model is universal and applicable to any other situation { (like to the upscattering photon spectra and the distribution of capital gain due to investement)} where the 
similar nature of processes occurs. 
}

{
\subsection{A possible origin of the HBB Component}

We find  an observational evidence of the HBB around 20 keV which can be fitted by a $\sim$ 4.5 keV blackbody  profile [see for example, Figures~\ref{sp_compar_cyg_refl_bbody}, \ref{sp_compar_gx_refl_bbody2},    
\ref{sp_evol_gx_LHS-IS}, \ref{sp_compar_grs_refl_bbody}, \ref{sp_all_bbody}  and Tables~\ref{tab:fit_grs}$-$\ref{tab:fit_4641_rxte}]. 
One can argue that this observable hump at 20 keV is a signature of the Compton reflection hump.
The reflection or downscattering  interpretation encounters a principal difficulty 
 because   the hard power-law tails of these spectra 
are too steep to form the Compton hump.
(see LT07). 
As it is seen from  Tables \ref{tab:fit_10512_cyg_x-1}-\ref{tab:fit_60091_cyg_x-1},  \ref{tab:fit_70109_gx}- 
\ref{tab:fit_4641_rxte} 
that   in almost all spectra where we found a noticable HBB  the photon index
$\Gamma > 2$.

{One can concern  regarding a way  how the current (reflection) paradigm  is excluded as the origin of the hump.   As we have already emphasize in the Intoduction that LT07 demonstrate  
that the characteristic downscattering hump is only formed when the spectral index, $\alpha<1$ (or $\Gamma<2$). 
It can be another counter argument against our interpretation of the hump that we use too simple models for fitting of the data not like  in more recent modeling [see e.g. the  relxill (reflionx) model] and the comparison with these more complex models (commonly used by the community in the context of the current paradigm). 
Now it becomes clear using our detailed investigations (see Fig.~\ref{time_scale_for_hbb_and_gaussian})  that the iron 
line component used in the reflection model  is not related  to the HBB  because they have very different time scales (at least two order difference in the time scales) and thus this modelling is not related to the real observations.

{
Specifically, Figure~\ref{time_scale_for_hbb_and_gaussian} demonstrates time-scale characteristics for HBB-feature ($red$) 
and Fe-Gaussian line ($blue$). Here we assume that if this  (iron-Gaussian) line is seen continuously in { some} 
observations (with approximately the same flux or parameters $N_{line}$, $E_{line}$), then this line is likely to 
remain visible in the  source spectrum  between these observations. As can be seen from our Tables, these interphases are smaller than the intervals of the observations themselves. In addition, we 
 { divided} the observational 
interval ($\sim$3 ks) into shorter sub-intervals. Thus we checked that the line is constant during the entire 
observation, i.e. it does not disappear and is not variable within this observational interval  
Then 
we summed up the times of all these intervals during a continuous sequence of observations, when this 
iron-Gaussian line is detected. 
But  for the HBB-feature case, we found observations (3 ks each), in which the HBB-feature is detected (see Table 3), and then 
we divide the entire observational interval into sub-intervals of 10--15 seconds, this HBB feature is not detected 
in some sub-intervals (Table 15). Then we chose a continuous chain  of sub-intervals, in which the HBB-feature 
is confidently detected and not detected before and after this chain  (we assume that HBB-feature is detected 
when its $EW \ge 70$ eV). Furthermore, we summed up the times of these sub-intervals (10--15 seconds each). Thus  we estimated the characteristic lifetime of the HBB-feature. In both cases (iron-Gaussian line and HBB feature), we  checked the stability of our result in a sample of all available observations.

As a result, we demonstrate the characteristic lifetime of the iron-Gaussian line in Figure~\ref{time_scale_for_hbb_and_gaussian} 
by red line against normalization, where the total time during which the iron-Gaussian line  continuously observed  is pointed  out  on the vertical axis. At the same time, along the horizontal axis, we indicated  normalizations ($N_{line}$)  
with which the iron-Gaussian line was observed.
The blue  points show HBB-feature 
detection: the vertical axis indicates the total time during which the HBB-feature was continuously observed 
(sometimes it was 3 ks, i.e. one observation, 800 s, i.e. less than the whole observations, and sometimes 
several successive observations). At the same time, along the horizontal axis, we indicate  normalizations 
($N_{HBB}$) with which the HBB-feature was observed (Table 15). This Figure is given for Cyg X--1, as an example, in which we wanted to show that the times of continuous visibility (we called this lifetime) of each of the features are very different. In the vertical axis, where the visibility time is set, the lifetime 
for the HBB-feature is $10^3-10^4$ s, while the lifetime for the Fe-line is $\ge 10^5$ s. In this Figure, we 
demonstrated that these spectral details are formed in geometrically  different areas of the source. 
Indeed, if HBB-feature and iron-line have different lifetimes, then most likely they are formed in different 
places of the source. Moreover, the HBB-feature is in a more compact area, i.e. closer to a BH, while the Fe-line is located in a wider area, i.e. farther from a BH. And since they are formed in completely different places, 
then most likely they have different conditions of formation and different origins. 

}

We describe  the HBB  hump  by a blackbody shape. This model has only two parameters: the color temperature and  normalization. The equivalent width (EW) of the hump should  be independent of the iron and other heavy elements, ionization parameter, because this hump is always distributed between 10 and 40 keV (see e.g. Fig. \ref{sp_all_bbody}) where there is no any of these lines. 
We should also emphasize that our spectral  model {\tt tbabs*(comptb+bbody)*smedge} plus one or two {\tt Gaussians} has only five free parameters.  It is worth noting that  we use seven parameters when we need to use two more parameters for the bbody component (HBB).
One can  advice us  to include the absorption edge in our best-fit models for {\it RXTE}, INTEGRAL, 
{\it Beppo}SAX data.  But  these  data do not require any additional absorption  components (see values of $\chi^2_{red}$ of the best-fits using our models in Tables \ref{tab:fit_10512_cyg_x-1}-\ref{tab:fit_60091_cyg_x-1},  \ref{tab:fit_70109_gx}-\ref{tab:fit_4641_rxte}.

}

Evolutions of spectral characteristics of Cyg X-1, GX 339-4, GRS 1915+105, SS~433 and V4641 Sgr  are very similar. While they have different BH masses
all of them  reveal the  HBB feature which is  centered around   20 
keV. 
{It is interesting   that the HBB feature is usually observed in the same 
X-ray luminosity range ($5\times 10^{36}$ -- $5\times 10^{37}$ erg/s) for different sources. 

One  can raise a natural question on the origin of the HBB. 
LT18  estimated the optical depth for photon-photon interaction  very close to a BH horizon. In order to do this  they  calculated the photon density $N_{\gamma}$ near a black hole horizon, assuming that the most photons there have energy greater than $m_ec^2=511$ keV:  
\begin{equation}
N_{\gamma}=\frac{L_{\gamma}}{4 \pi r^2 c m_e c^2},
\label{phot_dens}
\end{equation}  
where $L_{\gamma}\simeq 10^{37}(M_{bh}/10 M_{\odot})$ erg/s, $r = R_{sch} = 2GM/c^2$ is
the Schwarzschild radius, $R_{sch} = 3 \times 10^6(M_{bh}/10 M_{\odot})$ cm, 
$c$ is the speed of light ($3\times 10^{10}$ cm/s) and 
the electron rest energy, $m_e c^2$ is about $5\times 10^{-7}$ erg. As a result, LT18 obtained that $N_{\gamma}=0.6\times 10^{19}/(M_{bh}/10 M_{\odot})$ cm$^{-3}$. 

Then, LT18 estimate the optical depth for photon-photon interactions  as 
\begin{equation}
\tau_{\gamma - \gamma} \sim \sigma_{\gamma - \gamma} N_{\gamma} R_{sch}.
\label{tau_phot_phot}
\end{equation}  
The cross section for  $\gamma-\gamma$ interaction,  $\sigma_{\gamma - \gamma}\sim 0.2\sigma_T$, where $\sigma_T=6\times 10^{-25}$ cm$^2$ is 
the Thomson's cross-section. 
As a result one can find that    $\tau_{\gamma - \gamma} \sim (1 - 2)$ using 
Eqs. (\ref{phot_dens}--\ref{tau_phot_phot}).  It is worth noting that $\tau_{\gamma - \gamma}$ is independent of a BH mass.  

As  one can see the IS luminosity provides the sufficient $\tau_{\gamma - \gamma}$
   for the photon-photon interaction very close to a BH horizon.  On the other hand, pairs (electrons and positrons) are effectively  generated as a result of $\gamma - \gamma$ interaction
near a BH horizon. Some of the generated positrons propogate ourwards and in the way out they interacts with accreting electrons leading to the formation of the annihilation line, 511 keV. A significant fraction of these line photons can directly escape to the Earth observer if the Klein-Nishina optical depth, $\tau_{\rm KN}$ at 511 keV is of order of one. In the way out these 511 keV line photons undergo gravititational redshift with $z$ around 20 forming the HBB of the color temperature of around 5 keV.  Observationally this HBB feature has an equivalent width in the interval  from 400 to 800 eV (see Fig. \ref{EW_new+SS433}).
However, if the resulting  X-ray luminosities [$L_{\gamma} \gg  10^{37}(M_{bh}/10 M_{\odot})$ erg/s], then the dimensionless mass accretion rate  is much greater than 1  (see Titarchuk \& Zannias 1998) and pairs along with  the 511 keV annihilitation photons generated near a BH horizon  cannot escape because they are effectively scattered off converging electrons.
This our conclusion is firmly confirmed by the Monte Carlo simulations (see LT18). 
It is not  by chance that  we cannot see the HBB hump in the HSS. 
In the LHS we cannot see  the HBB and consequently  the redshifted annihilation line either  because 
 $\tau_{\gamma - \gamma} \ll 1 $ and pairs are not effectively generated and the converging flow is surrounded by the hot, relatively thick  Compton Cloud (CC) of the plasma temperature  of order of 50-60 keV.  The escaping  HBB photons are scattered off hot electrons of the CC.
}

\section{Discussion \label{discussion}}
One can ask various questions regarding the content of the presented  paper. For example, if he/she  takes a simple  exponentially cutoff power-law, some of which comes directly to the Earth observer and some of which shines on a cold disk and then comes to the Earth, what does that look like? In fact, this is an interesting question. If the index of this  exponentially cutoff power-law is too large then it should have high residuals at low-energies which is obvious. This is well-known effect of this simplification. As for the comparisons to standard reflection models one should be very careful to use it, because of two reasons: i. there  is no efficiency of the reflection for the spectral index $\alpha$  higher than 1 (or the photon index $\Gamma>2$ higher than 2).  In this case there is no a so called reflection hump. This is a very  strong theorem of the radiative transfer (see this proof in Laurent \& Titarchuk 2007). ii.  Another one  is that the disk illumination fraction of the photon emitted by the central object (COMPTON cloud) should be   only  about 25\%
(see details in Lapidus et al. 1985).

The next question is about  an observation  of a { reflection hump}.  Do we expect to see feature like those in the relxil code (i.e., a fluorescent Fe line and associated Fe edge,
fluorescent emission and edges at lower energies, an overall reduction in
photon energy, whether this manifests as changes in powerlaw slope and
cutoff energy, or also produces a { hump} for given photon indices? 
In fact, we use the Gaussian model  to describe a fluorescent Fe line and associated Fe edge in an additive manner since it is obvious that the iron line and its profile can be formed in different places of the source. In  Figure 5 we demonstrate  that there are {drastically} different time scales (at least two order of magnitude difference) between Fe emission line and HBB   component. This is  a very important observational result! 

 Does the answer that  the relxil code provides us  to the above questions are
fundamentally wrong? Or is the relxil code plausibly correct in the
spectral calculation that it  performs?  
The Relxill model describes both the continuum and emission features, without attracting additive components.  But for the final description of the spectra "in the relxill model", power and blackbody components are still involved, the interaction with which is not specified in terms of physical parameters.
 But what we have revealed  that the  different components of the spectra  of a source coming to the Earth  are formed in different parts of the source. At this stage using  the Relxill model  we are faced with a problem of the constancy of the photon index when the spectral state changes, which is impossible in terms of the spectral evolution established  by many authors using a huge set of BHs (see the Introduction). 
Usually in BHs the evolution of the photon index from 1.4 (LHS) through 2 (IS) to 3 (HSS) is observed. 

For example, Tomsick et al. (2018), in the end, advocate for a more
complex model in Cyg X-1 that  there is a directly viewed powerlaw, in
addition to the reflected powerlaw. But the reflection is relativistically
smeared in terms of Tomsick et al..  Although this model
has now become a much more complex model, it is not necessarily
unrealistic, since we know from Chandra-HETG data that there is a narrow component to the Fe line, likely coming from much further away from the innner accretion flow region because 
 Fe-line formed in the inner-most  parts of the accretion  flow could not survive.   Moreover this emission occurred there to be distorted due to  a coronal wind (see Laming  \&  Titarchuk 2004, Laurent \& Titarchuk 2007). 
 It is much easier to allow the formation of  iron lines  in the outer parts of the disk or in the wind. 
One can see the confirmation of this line formation in the extended part of the source (see 
Fig. \ref{time_scale_for_hbb_and_gaussian}).
Thus the  most simple version of the reflection models are
not particularly good descriptions of the data, as we show here.  

One can say that our model still
has issues that need to be addressed as a self-consistent
description of the iron line and smedge components. However, Laming \& Titarchuk (2004) and Laurent \& Titarchuk (2007) use the self-consistent models to describe the  iron-line and smedge components. Also one should keep in mind that the reflection effect does not take place if the spectral index $\alpha$ (or the photon index $\Gamma$) are very close to 1 (or 2) correspondingly.
Moreover, if $\Gamma>2$ for which we found a lot of data the reflection does not work at all while in the observations we  see the HBB hump  for these values of the photon index $\Gamma$.



\section{Conclusions \label{conclusions}}

{ In this paper we demonstrated the  HBB feature in the observed spectra of five  Galactical BHs and  suggest a scenario which allows us to understand how  this spectral feature is formed in a BH (see more details in LT18).  This HBB feature is usually   observed  in the IS which 
is  characterized by suitable conditions for the HBB observations, due to 
the low Klein-Nishina optical depth at  sufficiently high mass accretion rate of order 1--2, see LT18. 
In the IS $\Gamma \gax 2$  and thus accordingly to LT07 
the HBB hump  cannot be due to the  reflection effect. But in the LHS ($\Gamma \lax 1.5$),  
this hump hump can be a result of the downcattering  in an outflow (or reflection from the surrounding disk). 

Previously, we  reported a detection of the HBB component in the energy spectra around 20 keV in the GBHs, GRS~1915+105 and  SS~433 (see TS09, ST10).  
Now, we have presented a further study of this spectral feature using five  galactic BHCs: Cyg X-1, GX 339-4, GRS~1915+105, 
SS~433 and V4641~Sgr  using  our spectral models in  which Comptonization effects
are taken into account}. 
In particular, we have studied an evolution 
of X-ray energy spectra with  a short time step along the IS  of GRS~1915+105 as a sample of a bright X-ray source.  It was shown that the best-fit spectra obtained applying  a short  time exposition (10 -- 30 sec)  
indicate   to the HBB around  20 keV.

{ We detected the prominent HBB bump in Cyg~X--1 and GX 339--4 mainly in their IS and sometimes in LHS--IS ($1.8\le\Gamma\le 2$) and IS--HSS ($2\le\Gamma\le 2.5$) states, when the electron temperature is relatively low (see LT99, LT11, LT18).
The HBB  hump  increases when a BH  approaches to a classical IS ($\Gamma\sim 2.3$). 
This conclusion is based on extensive investigations  of all aforementioned BHCs.

We also presented spectral analysis of the INTEGRAL+{\it RXTE} observation which revealed the same HBB feature in SS~433. Furthermore, we  detected this HBB  in another Galactic BH, V4641~Sgr using the {\it RXTE} and $Beppo$SAX NFI detectors. 
We showed that the broadband energy spectra of { Cyg~X--1, GX~339--4}, GRS~1915+105, SS~433 and V4641~Sgr during all
observations could be adequately presented by  a sum of  the {Comptonization (BMC or COMPTB)} component,
the iron (Gaussian) line.  But  in quite a few observations of the   IS observations we should  add   the HBB component of   the color temperature $\sim$ 4.5 -- 5.9 keV to fit the data

We should also note that all these spectra with the HBB component are observed 
during the moderate radio active phase. 
Our observational results on the HBB hump  suggest that we observe the  
gravitationally redshifted annihilation line emission in these source (see details in LT18).
The detection of the HBB hump  as a redshifted annihilation line 
emission is possibly a new  observational evidence for the presence of a BH in these binaries. 

{
The detection of the redshifted  annihillation humps 
(which energy $E_{anh}=511$ keV in the comoving frame) in BH spectra at the energy from 10 to 40 keV
in the the lab frame} is a sensitive probe of the gravitational field in the Galactic black holes (Cyg~X--1, GX~339--4, GRS~1915+105, SS~433, and V4641~Sgr). This spectral feature is formed near the event horizon radius (at $R_S$) and thus this detection is a direct probe 
of the General Relativity effect.
}


\begin{acknowledgements}
We acknowledge the interesting remarks and points of the referee.
ES thanks  the IT support team (CNRS, LAPTh and LAPP, 
Annecy, France) for providing computing facilities.
ES also thanks the LAPTh, Annecy  (France) for hospitality and financial support of this work.

\end{acknowledgements}


\newpage
%
%

\begin{figure}[ptbptbptb]
\includegraphics[width=12cm]{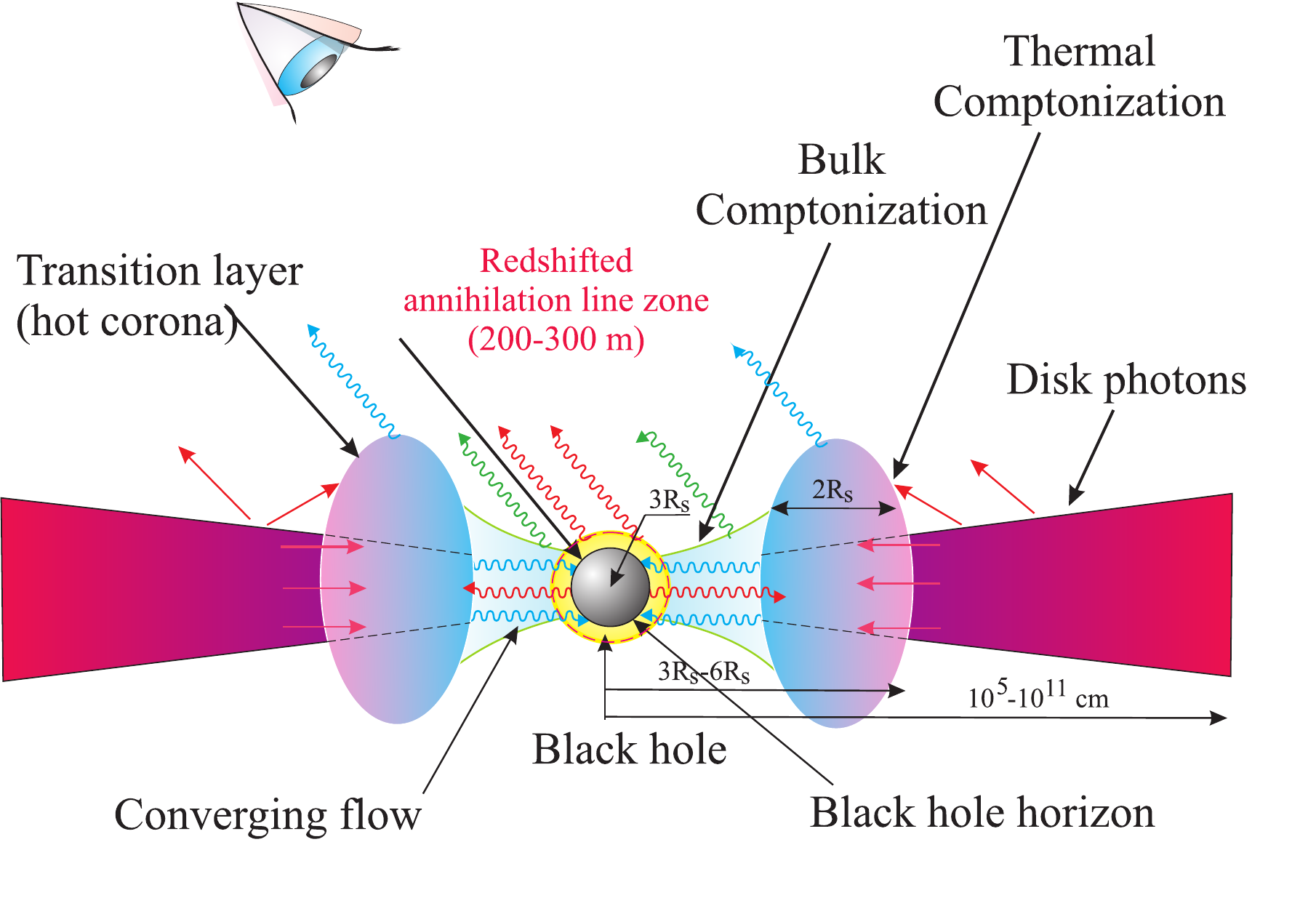}
\caption{A  proposed geometry for the thermal and bulk 
Comptonization regions in the binary hosting a BH.
The thermal plus 
bulk Comptonization spectrum 
(green and blue  arrrows) arises 
in the Compton cloud, where the 
disk BB seed photons undergo the thermal and dynamical Comptonization by  infalling 
material. 
In addition, 
we take into account the region near a BH horizon, where the photon-photon interactions lead to  the 
pair production effect. In this process the 
created positrons interact with accreting electrons there and therefore the annihilation line 
photons are created and distributed over the relatively narrow shell (enlarged in this plot) near 
the BH horizon. An Earth observer should see this annihilation line only at the gravitationally 
redshifted energies of the redshift  of $z\gg$1. 
}
\label{geometry}
 \end{figure}

%
%


\newpage
%
%

\begin{figure}[ptbptbptb]
\includegraphics[width=12cm]{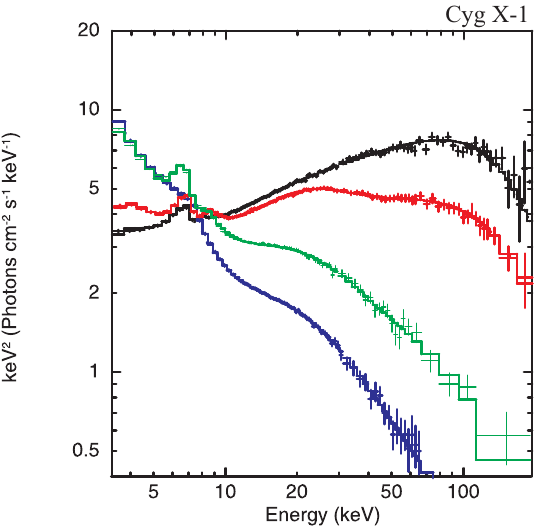}
\caption{Four representative $EF_E$ spectral diagrams during LHS, IS, HSS and VHS
spectral states of Cyg~X--1. Data are taken from {\it RXTE} observations
40417-01-03-00 (black, LHS), 90127-01-01-08 (red, IS), 60090-01-02-00 (green,
HSS), and 70414-01-01-02 (blue, VHS).}
\label{four_sp_cyg}
 \end{figure}

%
%

\newpage
\begin{figure}[ptbptbptb]
\includegraphics[width=16cm]{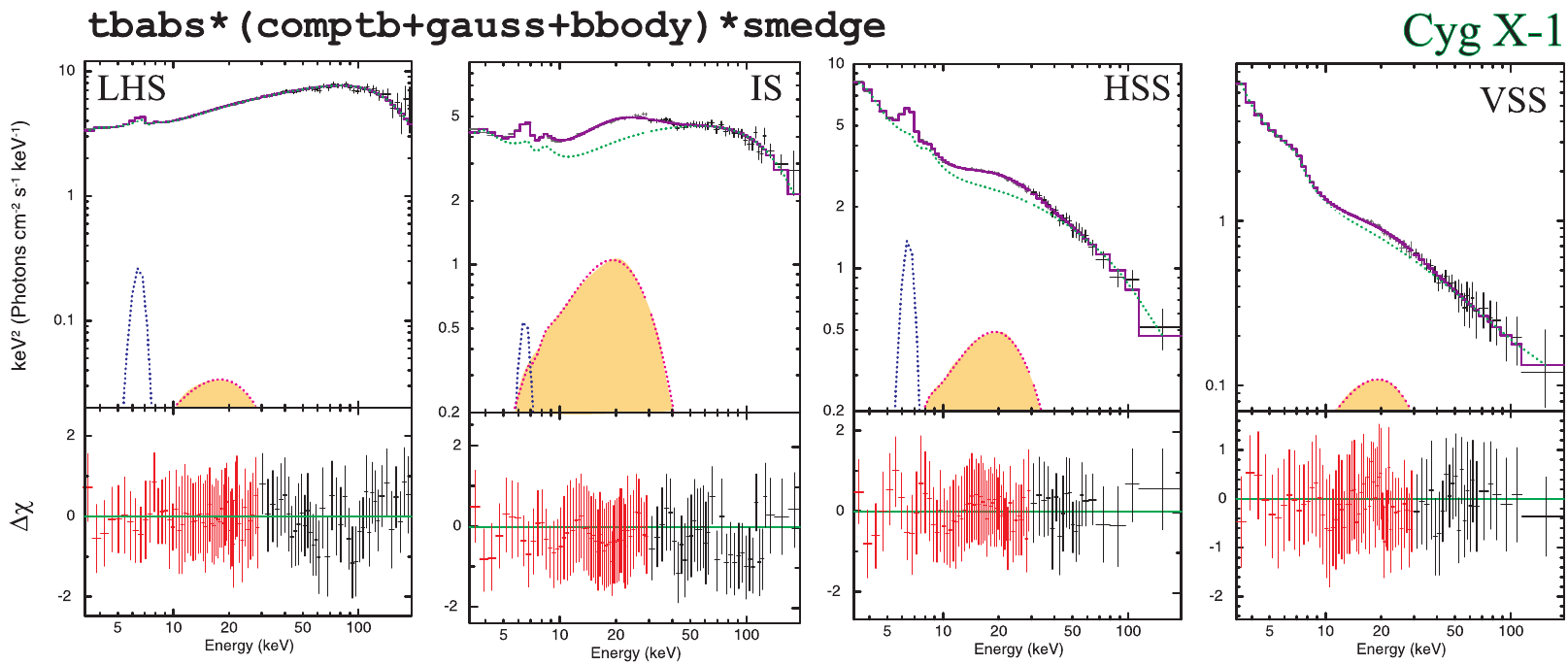}
\caption{
Evolution of  to Cyg~X--1 spectra using 
the {\tt tbabs*(comptb+gauss+bbody)*smedge} model 
for four spectral states. Data are taken the {\it RXTE} observations: 40417-01-03-00 (LHS, left), 90127-01-01-08 (IS, left center), 60090-01-02-00 (HSS, right center), and 70414-01-01-02 (VSS, right). 
The data are shown by black crosses and the spectral model components are displayed by dashed green, blue and pink lines for the {\tt comptb}, {\tt gaussian}, and {\tt Bbody}, respectively. {\it Yellow} shaded areas demonstrate an evolution of the HBB component during evolution between the LHS, IS, HSS, and VSS states.
}
\label{sp_compar_cyg_refl_bbody}
 \end{figure}

\newpage
%
%


\begin{figure}[ptbptbptb]
\includegraphics[width=14cm]{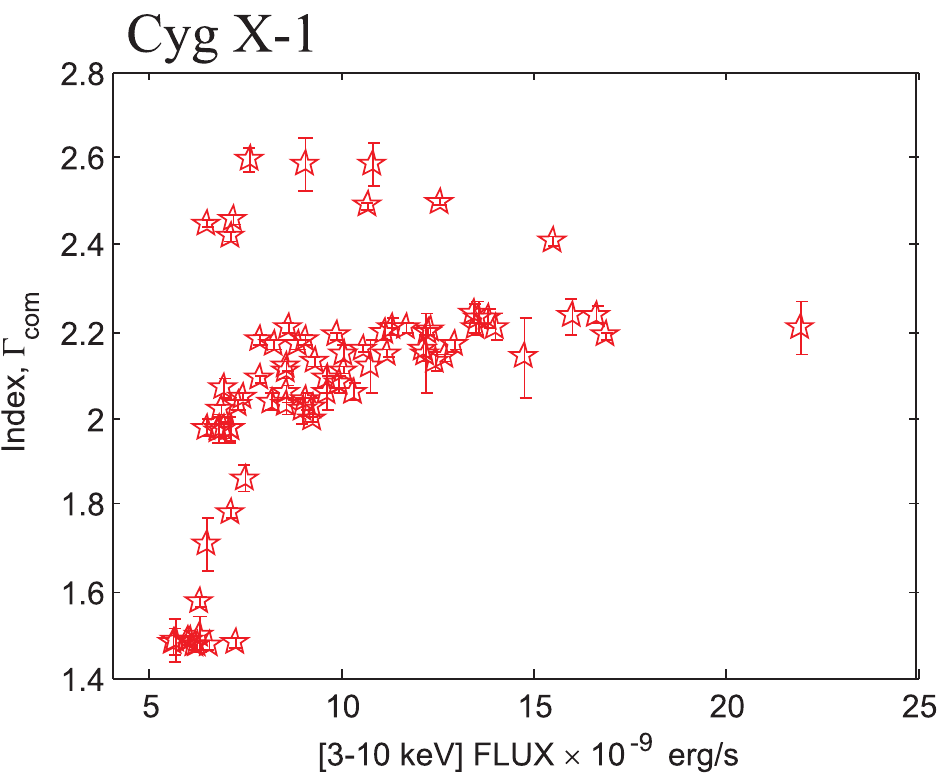}
 \centering
\caption{
The photon index versus the {\it RXTE} soft X-ray flux (3--10 keV) for Cyg~X--1 using the Comptb model (left panel) and  the reflection model (right panel).
}
\label{index_vs_soft_flux_comptb_refl_mod0}
 \end{figure}

\newpage

%
%


\begin{figure}[ptbptbptb]
\includegraphics[width=16cm]{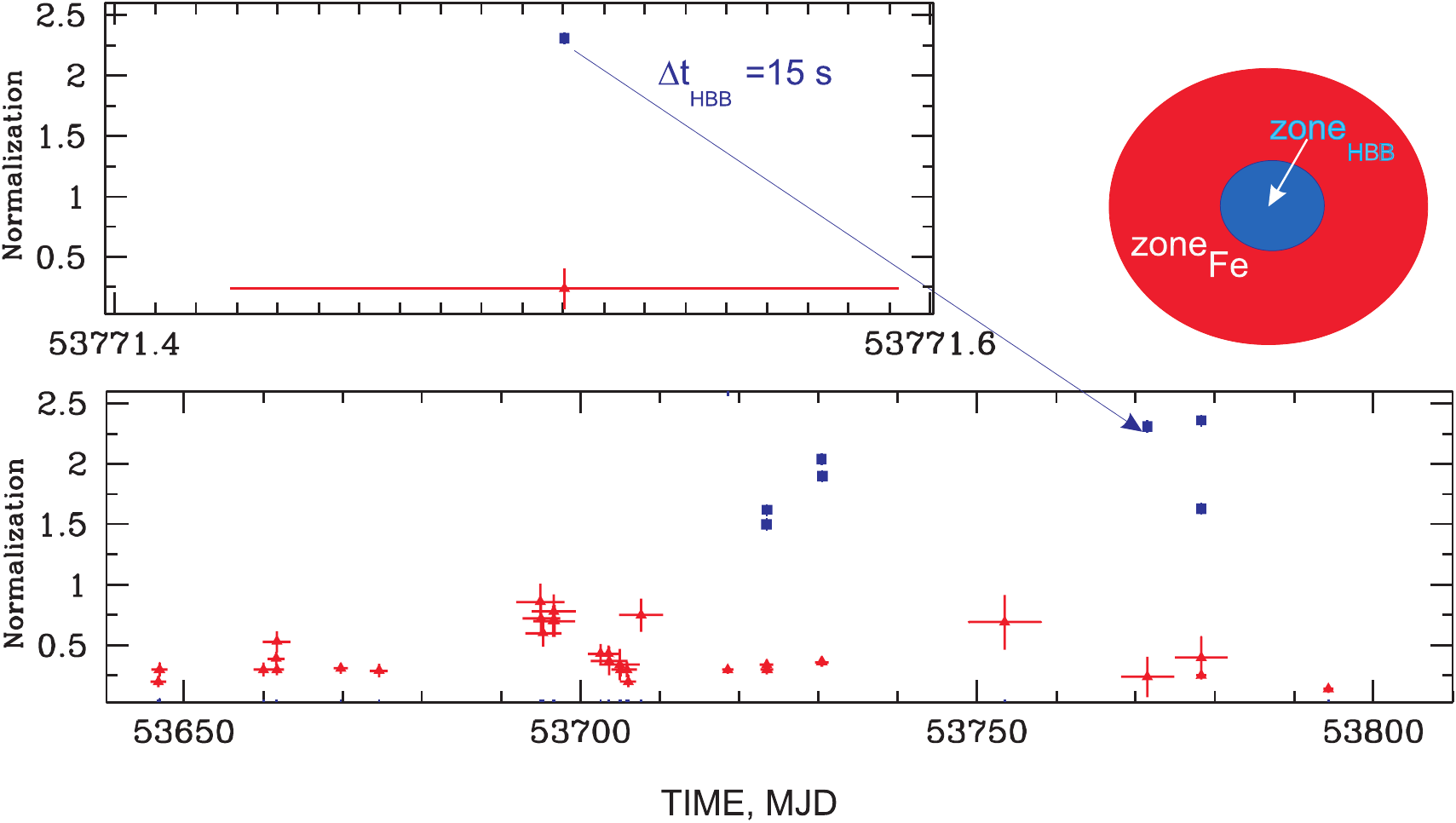}
 \centering
\caption{
Time scales for spectral features: the HBB component (red) and Fe emission line modified
by  absorption edge (blue), as a function of the feature normalization in $L_{39}/d_{10}^2$ 
(for the HBB) and total photons in units of cm$^{-2}$ s$^{-1}$ in the line for the Gaussian component  
(see Tables~6--9).
}
\label{time_scale_for_hbb_and_gaussian}
 \end{figure}

\newpage

%
%

\begin{figure}[ptbptbptb]
\includegraphics[width=12cm]{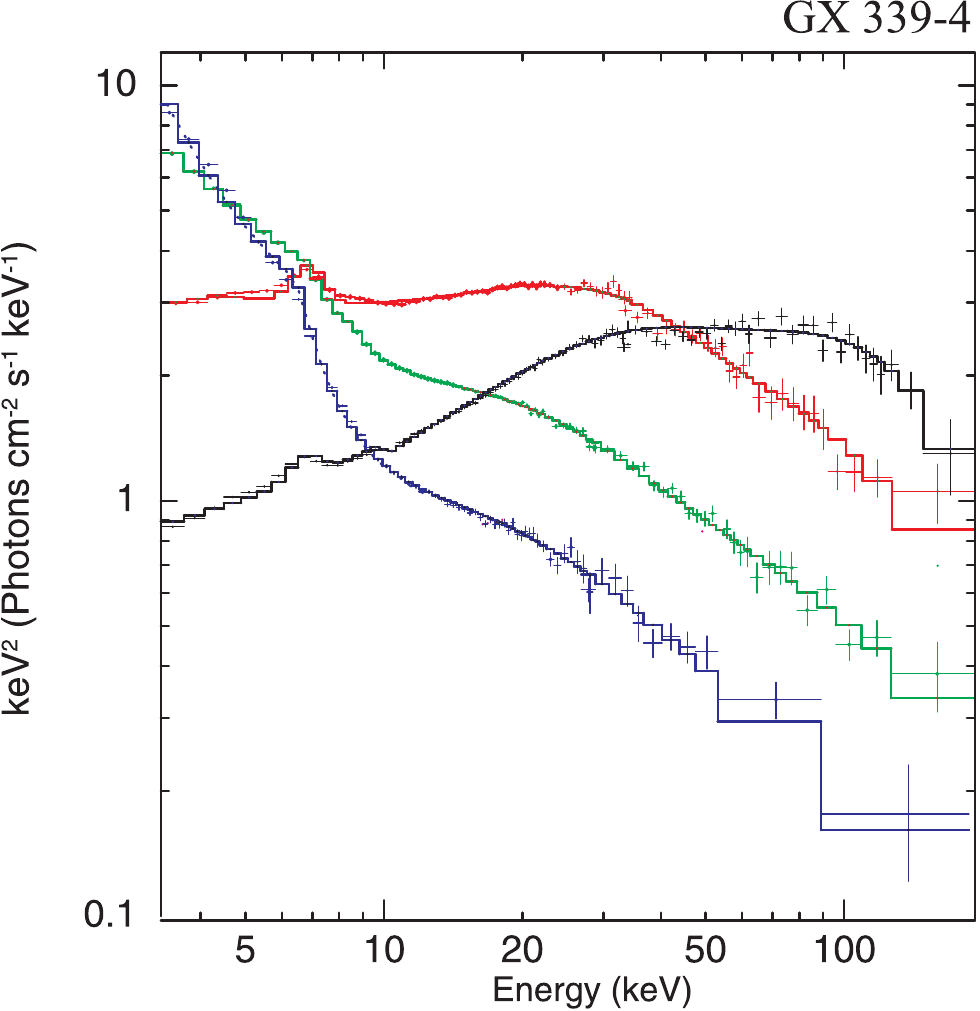}
\caption{ 
Four representative $EF_E$ spectral diagrams during the LHS, IS, HSS and VSS
spectral states of GX~339--4. Data are taken from {\it RXTE} observations
92428-01-01-00 (blue, LHS), 92035-01-03-00 (green, IS), 60705-01-69-01 (pink, IS), 70109-01-07-00 (red, HSS), and 70109-01-07-00 (black, VSS).}
\label{five_sp_gx}
 \end{figure}
\newpage

%
\begin{figure}[ptbptbptb]
\includegraphics[width=16cm]{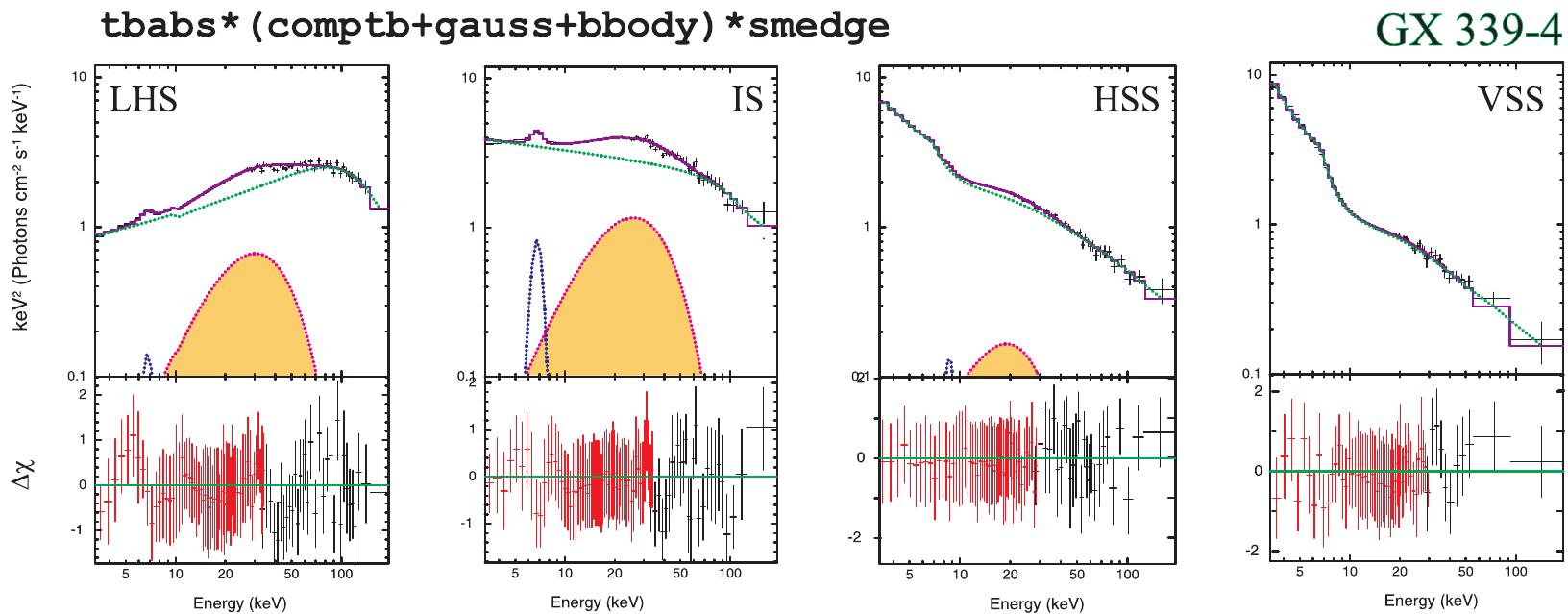}
\caption{Evolution of  to GX~339--4 spectra using 
the {\tt tbabs*(comptb+gauss+bbody)*smedge} model 
for four spectral states. Data are taken the {\it RXTE} observations 92428-01-01-00 (LHS-IS, left), 92035-01-03-00 (IS, left center), 70109-04-01-01 (HSS, right center), and 70109-01-07-00 (VSS, right). The colors and model component designations are the same as those in Fig.~\ref{sp_compar_cyg_refl_bbody}.
}
\label{sp_compar_gx_refl_bbody2}
 \end{figure}
\newpage

%
%

\begin{figure}[ptbptbptb]
\includegraphics[width=17cm]{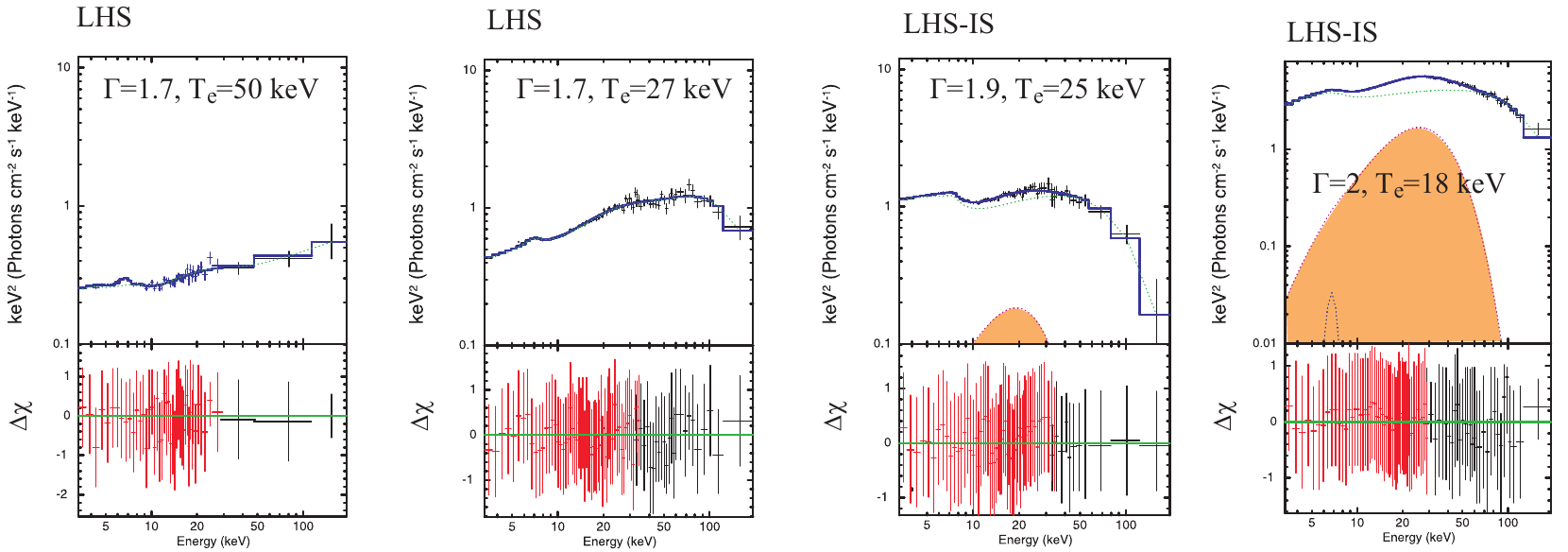}
\caption{Evolution of GX~339--4 spectra  between the LHS and IS states. 
Data are taken from the {\it RXTE} observations 70110-01-97-00 ($kT_e=50$ keV, $\Gamma$=1.7, LHS, left), 
90418-01-01-01 ($kT_e=27$ keV, $\Gamma$=1.7, LHS, left center), 60705-01-69-01 ($kT_e=25$ keV, $\Gamma$=1.9, 
LHS$\to$IS, right center), and 70110-01-08-00 ($kT_e=18$ keV, $\Gamma$=2.0, LHS$\to$IS, right). The colors and model component designations are the 
same as in Fig.~\ref{sp_compar_cyg_refl_bbody}.
{\it Yellow} shaded areas indicate the {\tt bbody} component, of which presence/absence can be associated with the the decreese/increase  
of the electron temperature $kT_e$  in the framework of {\tt tbabs*(comptb+gauss+bbody)*smedge} model. 
}
\label{sp_evol_gx_LHS-IS}
 \end{figure}

\newpage

%
%


\begin{figure}[ptbptbptb]
\includegraphics[width=12cm]{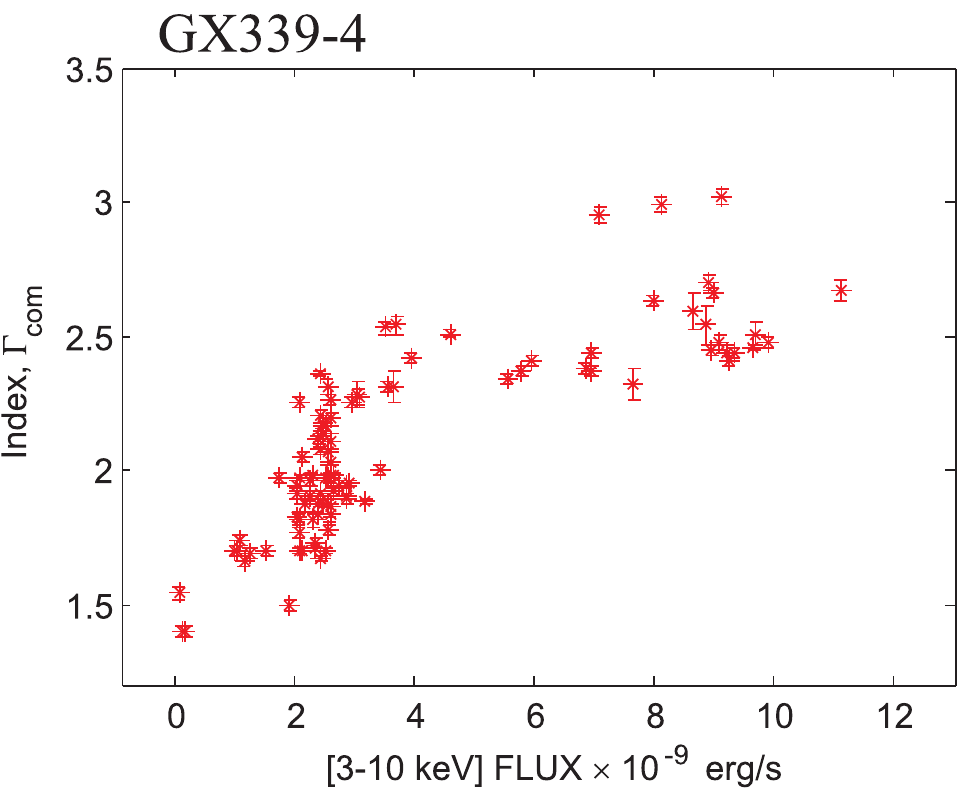}
 \centering
\caption{
The photon index versus the {\it RXTE} soft X-ray flux (3--10 keV) for GX~339--4 using the Comptb model (left panel) and  the reflection model (right panel).
}
\label{index_vs_soft_flux_comptb_refl_mod3}
 \end{figure}

\newpage

%
%

\begin{figure}[ptbptbptb]
\includegraphics[width=12cm]{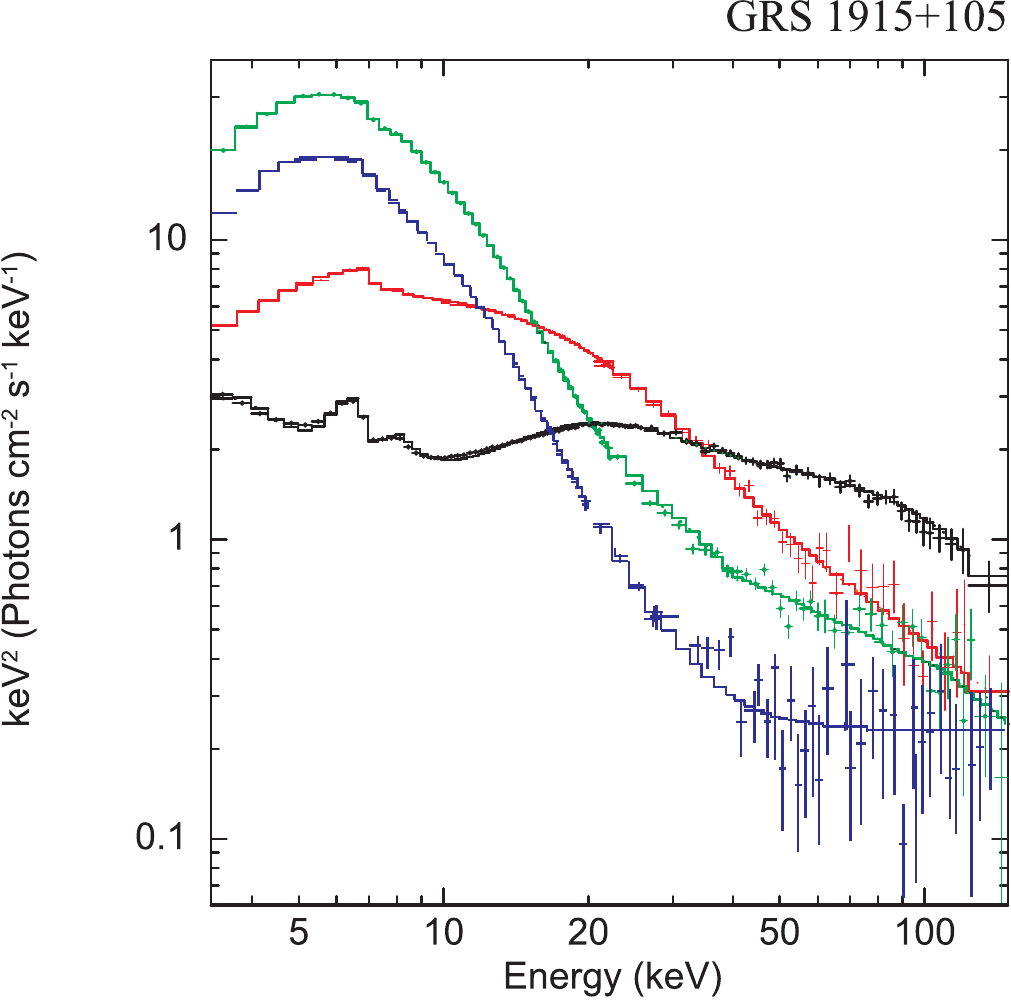}
\caption{
Four representative $EF_E$ spectral diagrams for the  LHS (black), IS (red),
HSS (green) and VSS (blue) for  GRS~1915+105. 
Data are taken using  {\it RXTE} observations.
}
\label{three_sp_grs}
 \end{figure}

\newpage

%
%

\begin{figure}[ptbptbptb]
\includegraphics[width=17cm]{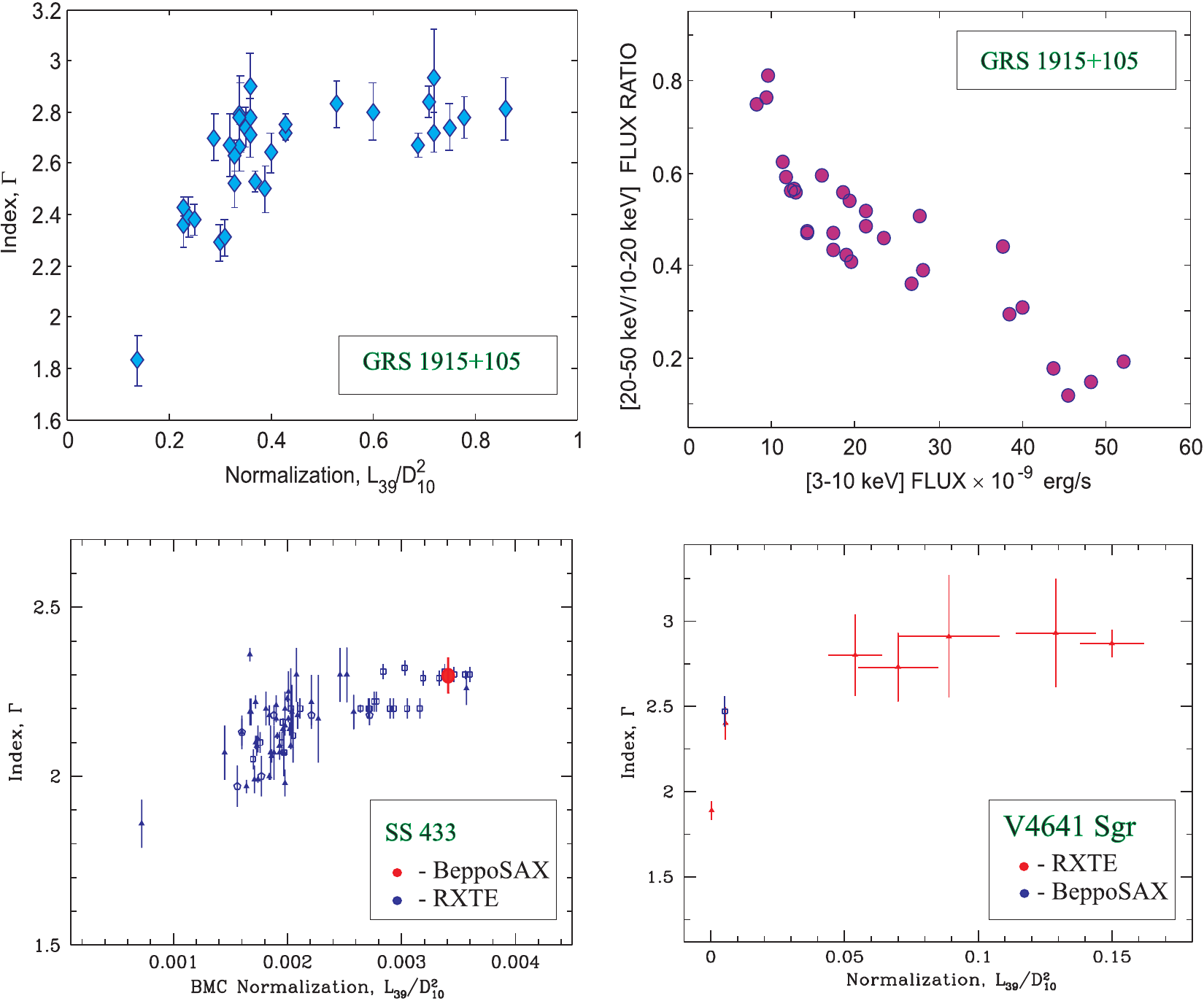}
\caption{Top:
{ The photon} 
index, $\Gamma$ plotted vs. Comptb normalization, propotional to the disk mass accretion rate (left) and 
hardness-flux diagram (right) for the 2005 -- 2006 transition of GRS~1915+105 analyzed using the tbabs*(comptb+gauss+bbody)*smedge model.
Bottom: The photon 
index, $\Gamma$ plotted vs. Comptb normalization for SS 433 (left) and for V4641 Sgr (right), respectively. 
}
\label{gam_2005}
\end{figure}

\newpage

%
%

\begin{figure}[ptbptbptb]
\includegraphics[width=16cm]{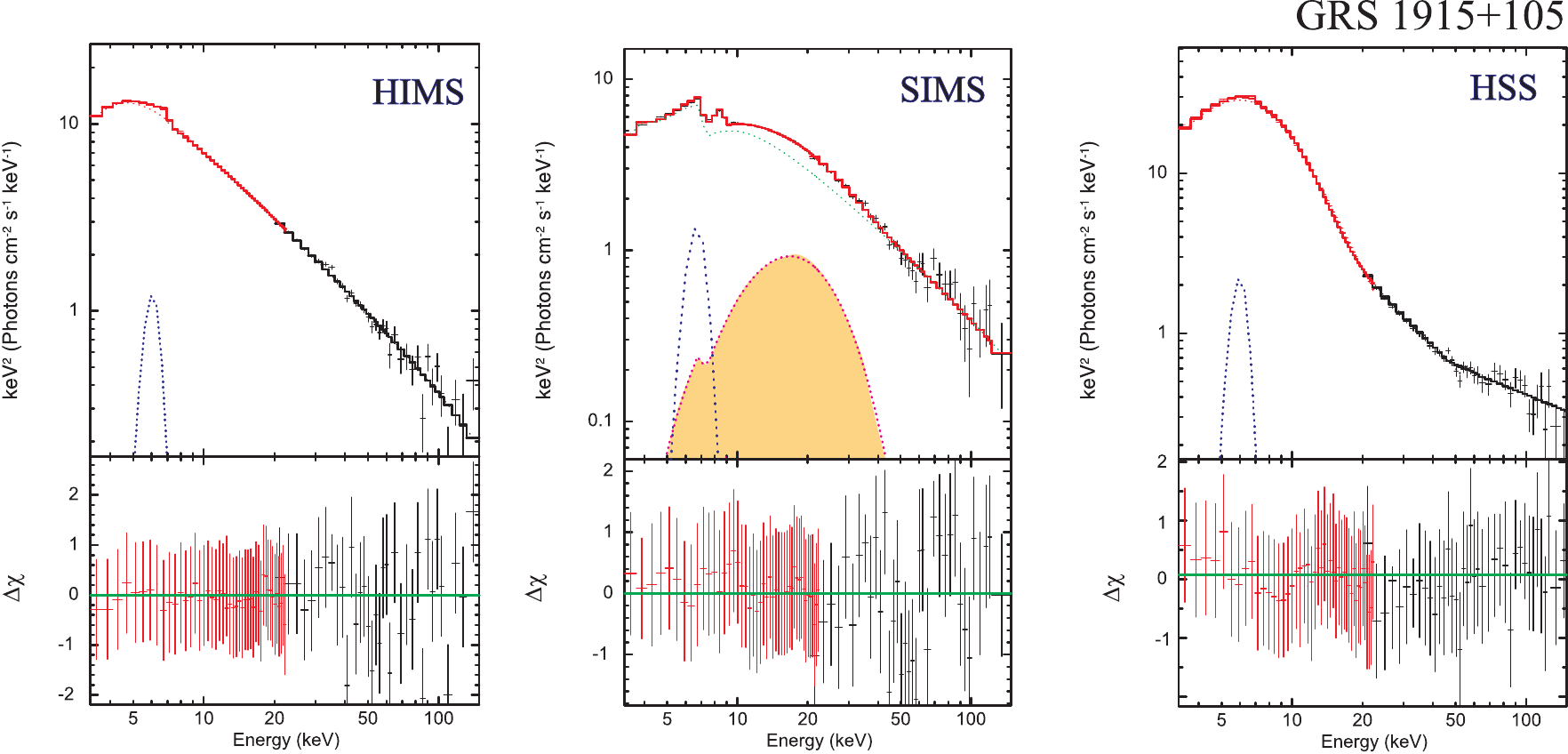}
\caption{
Examples of the best-fit spectra of GRS~1915+105 for three spectral states using {\tt tbabs*(comptb+gauss+bbody)*smedge} model. 
Data are taken from {\it RXTE} observations
90024-02-43-00 (left, HIMS), 91701-01-49-00 (center, SIMS) and 92092-02-01-00 (right, HSS).
}
\label{sp_compar_grs_refl_bbody}
 \end{figure}


\newpage

%
%

\begin{figure}[ptbptbptb]
\includegraphics[width=15cm]{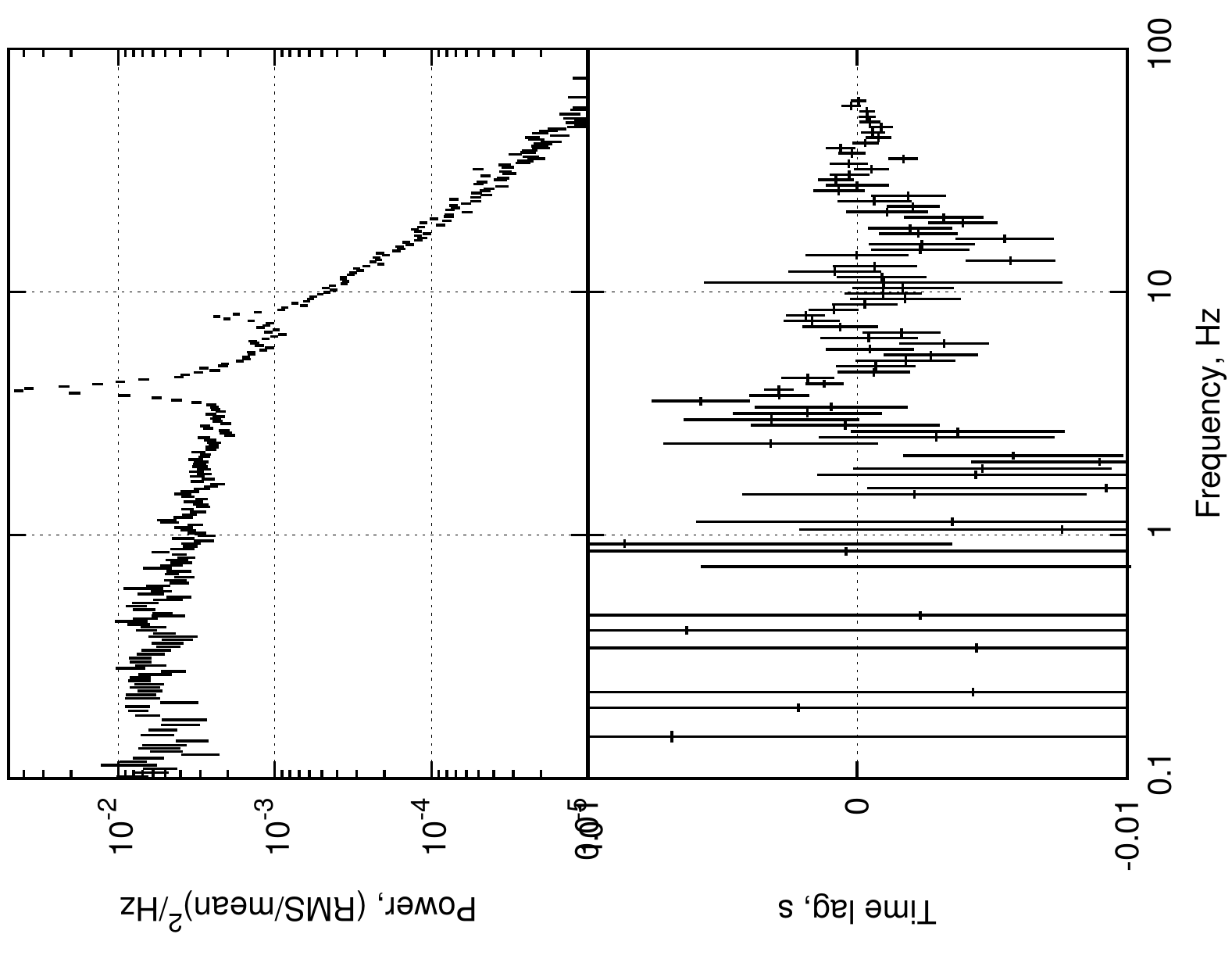}
\caption{{ From top to bottom:}
 evolutions of the flux density $S_{15GHz}$ at 15 GHz ({\it Ryle} Telescope), the {\it RXTE}/ASM count rate, 
 Comptb normalization, and the photon index, $\Gamma$ in the 2005 -- 2006  transitions 
of GRS 1915+105 (MJD 53690 -- 53855).
All vertical strips mark the time intervals, in which the HBB  
component  was detected. A behavior of spectral parameters for the  short time 
10 -- 30 s intervals (see Tables~\ref{tab:par_bbody}$-$\ref{tab:fit_grs_laor}) occurred  during  an interval of MJD=53771.51 -- 53772.1 
(pink vertical strip, 91701-01-49-00 observation). 
}
\label{lc_grs}
 \end{figure}


\newpage
%
%

\begin{figure}[ptbptbptb]
\includegraphics[width=15cm]{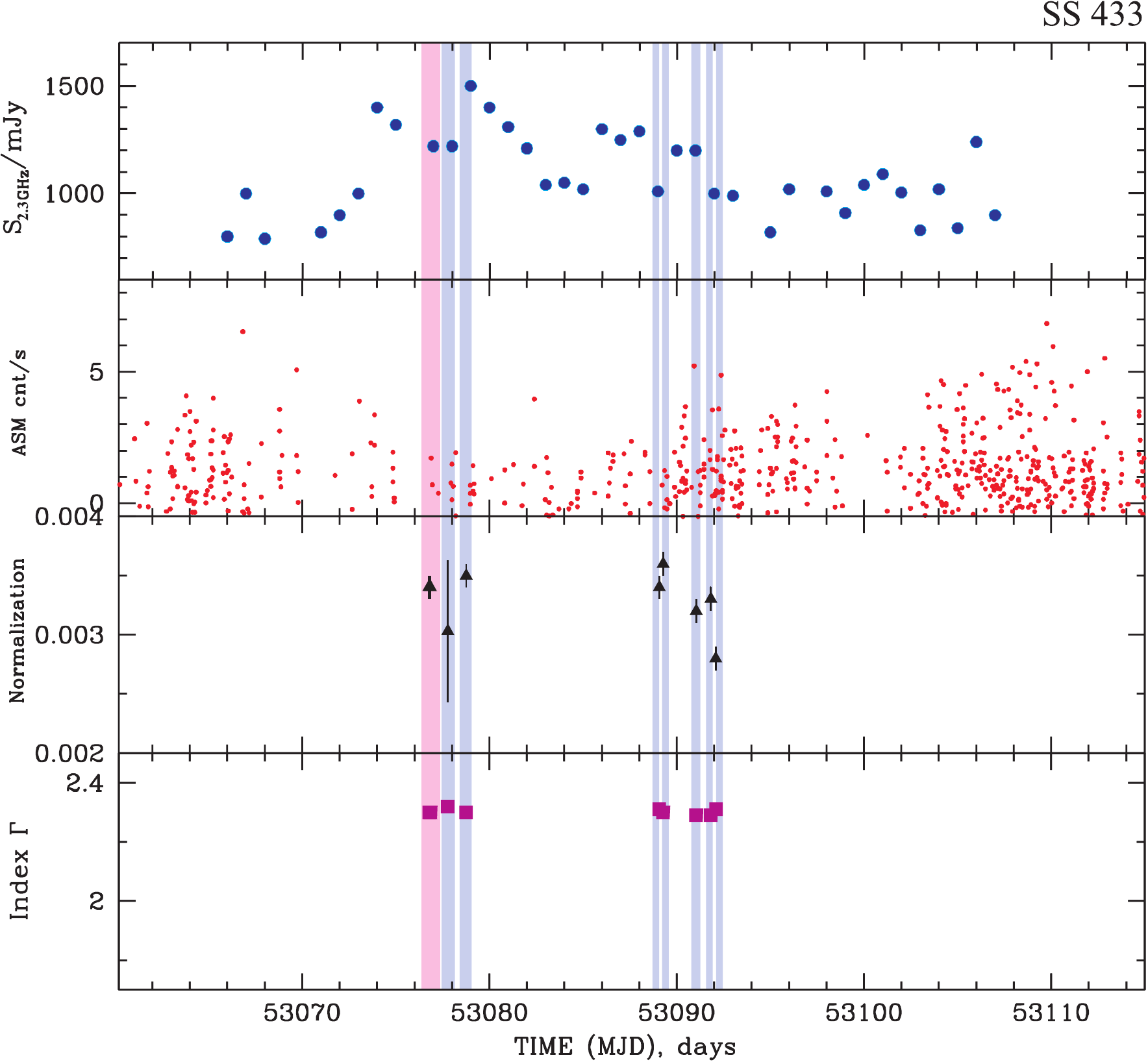}
\caption{{\it From top to bottom:}
 evolutions of the flux density, $S_{2.3GHz}$ at 2.3 GHz (RATAN-600), {\it RXTE}/ASM count rate, Comptb normalization, and the photon index, $\Gamma$ during March -- April 2004  transition of SS~433 (MJD 53060 -- 53114). 
Vertical strips mark the time intervals, in which an additional spectral 
component, HBB was detected. Pink vertical strip indicates the time 
interval of a simultaneous INTEGRAL and {\it RXTE} observation of SS~433. 
}
\label{lc_ss}
 \end{figure}


%
%

\newpage

%
%


\begin{figure}[ptbptbptb]
\includegraphics[width=14cm]{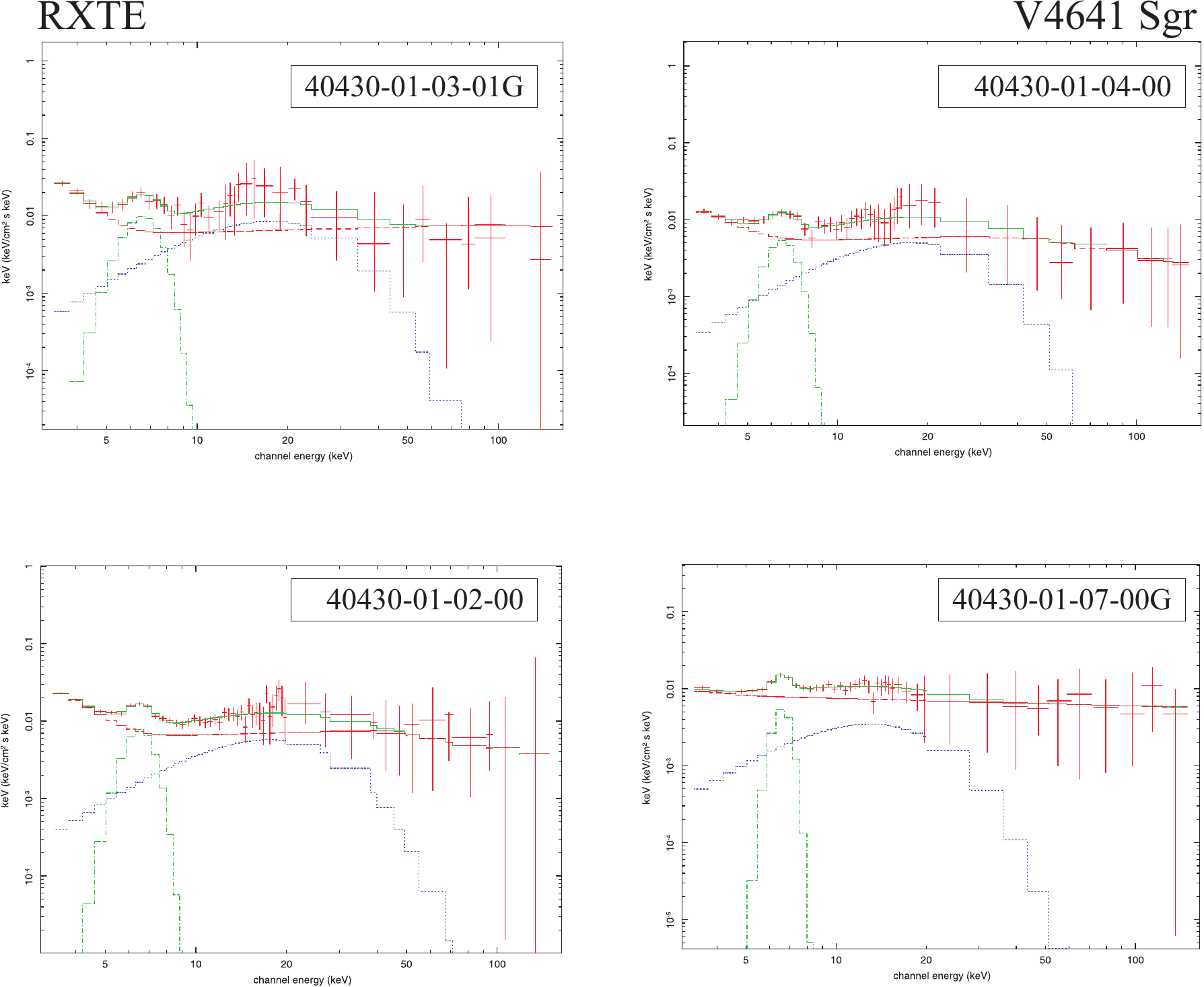}
\includegraphics[width=8cm]{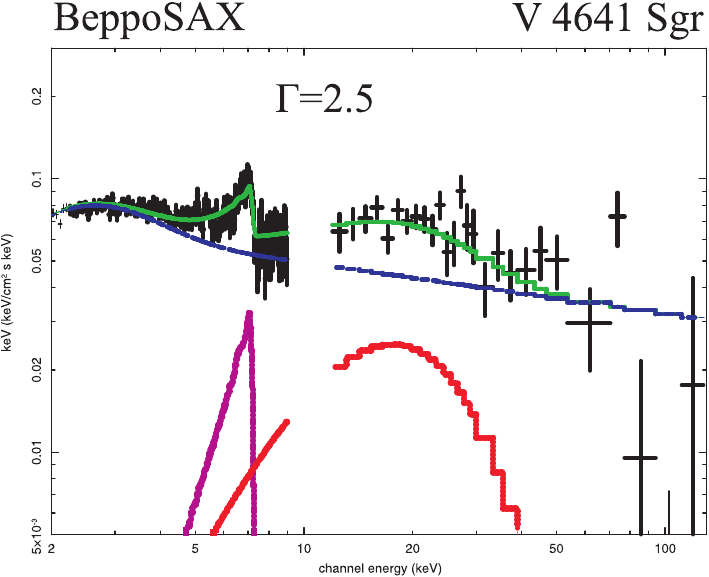}
 \centering
\caption{
We apply a model, {\tt tbabs*(comptb+gauss+bbody)*smedge} to the BH binary V4641 spectra (see Tables \ref{tab:fit_4641}$-$\ref{tab:fit_4641_rxte}) and detected the HBB in four {\it RXTE} spectra of V4641 Sgr: 40430-01-02-00,  40430-01-03-01G, 
40430-01-04-00, 40430-01-07-00G and in the {\it Beppo}SAX spectrum. 
Top: the data are shown by red crosses and the spectral model components are displayed by dashed red, green and blue lines for the Comptb,  gaussian, and HBB, respectively. Bottom: 
the data are shown by black crosses and the spectral model components are displayed by   blue,  pink and red lines for the  Comptb,   Laor, and HBB, respectively.
}
\label{sp_all_bbody}
 \end{figure}


\newpage

%
%

\begin{figure}[ptbptbptb]
 \centering
\includegraphics[width=14cm]{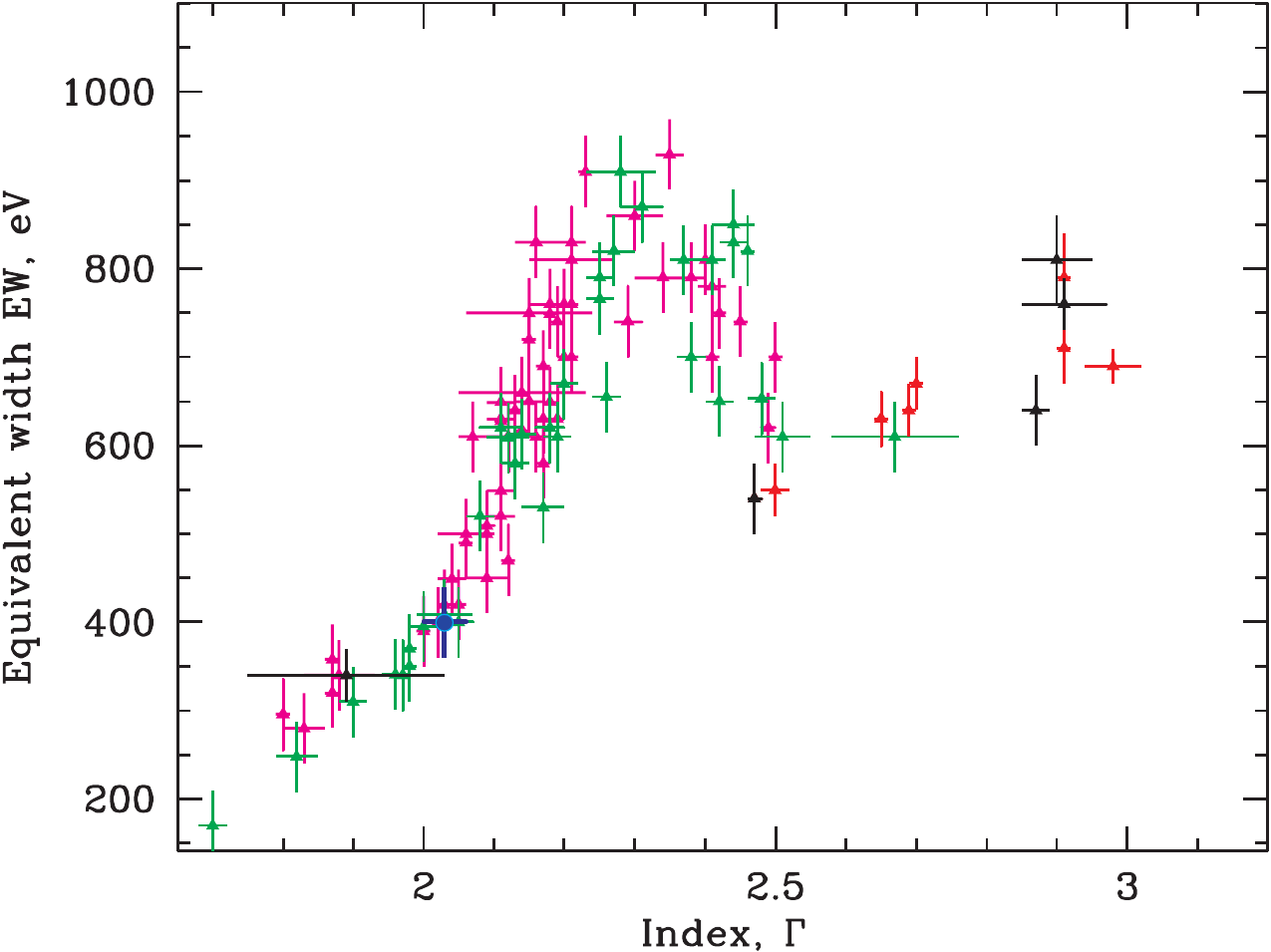}
\caption{
Equivalent width of the HBB in eV as a function of the  
photon index, $\Gamma$  {for different sources: Cyg X-1 (pink), GX 339-4 (green), SS~433 (blue), GRS~1915+105 (red) and V4641~Sgr (black)}.
}
\label{EW_new+SS433}
 \end{figure}




\newpage



                }

\end{document}